\documentclass[aps,prd,twocolumn,showpacs,superscriptaddress,nofootinbib]{revtex4-2}
\usepackage{dcolumn}
\usepackage{blindtext}
\usepackage{hyperref}
\usepackage{amsmath,amssymb}
\usepackage{float}
\usepackage{microtype}
\usepackage{graphicx}
\usepackage{bm}
\usepackage{latexsym}
\usepackage{epsfig}
\usepackage{psfrag}
\usepackage{color}
\usepackage[dvipsnames]{xcolor}
\usepackage{subfigure}
\usepackage[section]{placeins}
\usepackage{natbib}
\usepackage{ulem}
\usepackage{lipsum}
\usepackage{comment}

\def\l{\left}
\def\r{\right}

\def\nn{\nonumber}

\def\ns{n_{\rm s}}
\def\vx{\mathbf{x}}
\def\vk{\mathbf{k}}
\def\vq{\mathbf{q}}

\def\Mp{\mathrm{M}_{\rm P}}
\def\HI{H_{\rm e}}
\def\ke{k_{\rm e}}
\def\kre{k_{\rm re}}
\def\ee{\eta_{\rm e}}
\def\ere{\eta_{\rm re}}
\def\mH{\mathcal{H}}
\def\mB{\mathcal{B}}

\def\nb{n_{\rm B}}
\def\ns{n_{\rm s}}
\def\kpv{k_{*}}

\def\GeV{\text{GeV}}

\def\Nk{N_{\kpv}}
\def\Nre{N_{\rm re}}
\def\As{A_{s}}

\def\nc{n_\zeta}

\def\Tre{T_{\rm re}}
\def\wre{w_{\rm re}}
\def\ae{a_{\rm e}}
\def\are{a_{\rm re}}
\def\Mpc{\text{Mpc}}

\def\hk{h_{\vk}^{\lambda}}

\def\zetab{\zeta_{\rm B}}

\def\mPc{\mathcal{P}_{\zeta}}

\def\mPcphi{\mathcal{P}_{\zeta}^{\Phi}}
\def\mPcb{\mathcal{P}^{\delta \rho_{B}}_{\zeta}}

\def\mPb{\mathcal{P}_{\rm B}}
\def\tmPb{\tilde{\mathcal{P}}_{\rm B}}
\def\mPe{\mathcal{P}_{\rm E}}
\def\tmPe{\tilde{\mathcal{P}}_{\rm E}}
\def\mPt{\mathcal{P}_{\rm T}}

\def\mPts{\mathcal{P}_{\rm T}^{\rm sec}}

\def\rhob{\rho_{\rm B}}

\def\d{\mathrm{d}}

\def\mAk{\mathcal{A}_{\vk}}

\def\qnb{q_{\nb}}
\def\rhoem{\rho_{\rm EM}}

\def\mGk{\mathcal{G}_{k}}
\def\bJ{\mathrm{J}}
\def\tx{\tilde{x}}
\def\mF{\mathcal{F}}

\def\th{t_{\rm h}}
\def\Mpbh{M_{\rm PBH}}
\def\rhopbh{\rho_{\rm PBH}}
\def\rM{\rm M}
\def\fpbh{f_{\rm PBH}}
\def\Opbh{\Omega_{\rm PBH}}
\def\Odm{\Omega_{\rm DM}}
\def\gsre{g_{\rm re}^{*}}
\def\gseq{g^{*}_{\rm eq}}
\def\gsp{g^{*}_0}

\def\hij{h_{ij}}
\def\hkl{h_{\vk}^{\lambda}}
\def\hinf{h_{\vk}^{\rm inf}}
\def\hkr{h_{k,\rm ra}^{\lambda}}
\def\hkrp{h_{k,\rm ra}^{\lambda '}}
\def\xre{x_{\rm re}}
\def\ogwh{\Omega_{\rm gw}h^2}
\def\ogw{\Omega_{\rm gw}}
\def\ogwp{\Omega_{\rm gw}^{\rm pri}}
\def\ogws{\Omega_{\rm gw}^{\rm mag}}

\def\mJ{\mathrm{J}}
\def\mD{\mathcal{D}}
\def\fre{f_{\rm re}}
\def\fe{f_{\rm e}}
\def\xe{x_{\rm e}}
\def\mJ{\mathrm{J}}
\def\mAki{\mathcal{A}_{\vk}^{\mathrm{inf}}}

\def\ksb{k_{\rm SB1}}
\def\ksbs{k_{\rm SB2}}

\allowdisplaybreaks[1]

\begin{document}

\title{The Magnetic Origin of Primordial Black Holes: A Viable Dark Matter Scenario}
\author{Subhasis Maiti}
\email{E-mail: subhashish@iitg.ac.in}
\affiliation{Department of Physics, Indian Institute of Technology, Guwahati, 
Assam, India}
\author{Debaprasad Maity}
\email{E-mail: debu@iitg.ac.in}
\affiliation{Department of Physics, Indian Institute of Technology, Guwahati, 
Assam, India}

\begin{abstract}
The origin of primordial magnetic fields and the nature of dark matter remain open problems in cosmology, largely due to the absence of direct observational probes of early-Universe magnetogenesis. Primordial black holes (PBHs) provide a potential link between these two issues, as their formation is sensitive to small-scale energy-density fluctuations.
In this work, we investigate PBH formation sourced by primordial magnetic fields generated in the early Universe. We consider a magnetogenesis scenario that can account for the observed large-scale magnetic fields while also allowing PBH formation in a mass range consistent with PBHs constituting a significant fraction of the cold dark matter.
We further analyze the stochastic gravitational-wave background produced by magnetic-field–induced anisotropic stresses and show that, for certain regions of parameter space, the resulting signal lies within the projected sensitivity of future gravitational-wave observatories such as LISA, DECIGO, BBO, and SKA. By comparing the parameter dependence of the PBH abundance and the gravitational-wave spectrum, we demonstrate that these observables provide complementary constraints on the underlying magnetogenesis model.
Our results illustrate how combining PBH and gravitational-wave observations can improve our ability to test magnetogenesis scenarios and probe early-Universe dynamics.
\end{abstract}
\maketitle


\section{Introduction}
Primordial Black Holes (PBHs) are hypothetical black holes that may have formed in the early Universe, well before the emergence of stars and galaxies. They originate from the collapse of overdense regions re-entering the Hubble horizon during the post-inflationary era, seeded by fluctuations generated during inflation as well as post-inflationary dynamics. When the curvature perturbation amplitude exceeds a critical threshold, gravitational collapse leads to PBH formation, a mechanism first proposed by Zel’dovich and Novikov~\cite{Zeldovich:1967lct} and later developed by Hawking and Carr~\cite{Hawking:1971ei, hawkings, 1975ApJ...201....1C}.

Compared with the stellar backs holes, PBHs do not require any stellar progenitors and are instead governed purely by high-energy physics in the early Universe. This makes them unique probes of inflation, reheating, and beyond-standard-model phenomena at energy scales far beyond those accessible to particle accelerators. PBHs have drawn significant interest in recent years due to their potential roles as dark matter candidates~\cite{PhysRevD.50.4853, Carr:2003bj, Carr:2016drx, Sasaki:2016jop, Carr:2020xqk, Green:2020jor, Green:2024bam, Borah:2025wzl}, as seeds for supermassive black holes (SMBHs)~\cite{Bean:2002kx, PhysRevD.70.064015, Kawasaki:2012kn, Ziparo:2024nwh, Ziparo:2024nwh}, and as contributors to baryogenesis and cosmic radiation via Hawking evaporation~\cite{Majumdar:1995yr, Baumann:2007yr, Hook:2014mla, Smyth:2021lkn, Datta:2020bht, Boudon:2020qpo, Morrison:2018xla, PhysRevD.59.041301, Bernal:2022pue, Schmitz:2023pfy, Borah:2024lml, Barman:2022pdo, RiajulHaque:2023cqe, Hamada:2016jnq, DeLuca:2022bjs, DeLuca:2021oer}.

Importantly, PBH production is highly sensitive to the small-scale primordial curvature power spectrum, which remains unconstrained by CMB or large-scale structure observations~\cite{SDSS:2003eyi, Khatri:2013dha, Planck:2018jri, DES:2021wwk}. This opens up opportunities for mechanisms that enhance perturbations on these scales. Among the best-studied models are inflationary scenarios with an ultra-slow-roll (USR) phase~\cite{Garcia-Bellido:2017mdw, Motohashi:2017kbs, Byrnes:2018txb, Ballesteros:2018wlw, Raveendran:2022dtb, Ragavendra:2020sop, 
  Garcia-Bellido:1996mdl, Garcia-Bellido:2017mdw, PhysRevD.103.083510, Solbi:2021wbo, Figueroa:2021zah, Frolovsky:2022qpg, PhysRevLett.102.161101}, where the inflaton temporarily slows down on a flat potential, amplifying superhorizon fluctuations. Other possibilities include PBH formation via bubble collisions in first-order phase transitions~\cite{PhysRevD.109.123030, PhysRevD.26.2681, Rubin:2001yw, Ai:2024cka} or the collapse of topological defects like cosmic strings~\cite{Hawking:1987bn, Polnarev:1988dh, PhysRevD.45.3447, Balaji:2025tun, Balaji:2024rvo, Balaji:2022rsy}.

In this work, we propose yet another physically motivated alternative mechanism utilizing the well studied inflationary magnetogenesis model. We show that even without a USR phase, simple magnetic spectra with power law form generated during the early stage of our universe (inflation and during reheating), can enhance curvature perturbations sufficiently to produce PBHs. 
Most studies on magnetogenesis models focus on matching present-day magnetic field observations at large scales and associated gravitational wave production. 
However, their impact on primordial curvature perturbations has received comparatively less attention.
Interesting to note that the magnetic field at small scales is poorly constrained, similar to the inflationary scalar power spectrum. This freedom motivates us to tune the magnetic spectrum at small scales rather than tuning the inflatonary phase, and enhance the scalar power spectrum. By suitably tuning the magnetic spectrum, we indeed show that the scalar power spectrum can be enhanced sufficiently; therefore, neither an ultra-slow-roll (USR) phase nor other non-standard inflationary dynamics are required, yet a substantial PBH abundance can still be achieved.

In this work, we focus on a special class of magnetogenesis model (see~\cite{Maiti:2025awl} for details)
that is capable of simultaneously explaining the large-scale magnetic fields observed by various experiments and producing PBHs that can constitute a partial CDM component. Furthermore, for specific parameter choices, they can yield a significant PBH abundance in mass ranges where PBHs could serve as the sole dark matter candidate. We find that the PBH mass is primarily determined by the reheating temperature, while the PBH fractional energy density depends both on the reheating dynamics and the present-day magnetic field.

Magnetic fields are observed across a broad range of astrophysical environments, from kiloparsec (Kpc) to megaparsec (Mpc) length scales, with significant strength. Most intriguing is the presence of a magnetic field in the intergalactic medium (IGM), with length scales even larger than Mpc, and a lower bound on the present-day magnetic field strength around $10^{-17}-10^{-15}\,\text{G}$~\cite{2011A&A...529A.144T, PhysRevLett.116.191302, 2010Sci...328...73N, 2011A&A...529A.144T}. Furthermore, at similar scales, the CMB anisotropies yield an upper bound on the present-day magnetic field strength of order $\sim \text{nG}$~\cite{Planck:2015zrl, Paoletti:2008ck, BICEP2:2017lpa}. The origin of such large-scale magnetic fields remains one of the most challenging problems in cosmology and has been widely discussed in the literature. 

Inflationary magnetogenesis models~\cite{Guth:1980zm,Linde:1981mu, Albrecht:1982wi, Starobinsky:1980te, Kandus:2010nw, Durrer:2013pga, Ferreira:2013sqa, Subramanian:2015lua, Kobayashi:2014sga, Haque:2020bip, Tripathy:2021sfb, Li:2022yqb, Adshead:2016iae, Maiti:2025cbi,PhysRevD.94.043523, Sharma:2017eps,Ng:2014lyb, Papanikolaou:2024cwr, Hortua:2014wna, Maiti:2025rkn, Maiti:2025awl} are among the most popular scenarios. A typical primordial inflationary magnetogenesis setup is modeled by introducing couplings between the gauge field and scalar or pseudoscalar fields, e.g., $ f^2(\phi,R) F_{\mu\nu} F^{\mu\nu}$~\cite{Demozzi:2009fu, Ratra:1991bn, Kandus:2010nw, Durrer:2013pga, Ferreira:2013sqa, Subramanian:2015lua, Kobayashi:2014sga, Haque:2020bip, Tripathy:2021sfb, Li:2022yqb,Ng:2014lyb, Maiti:2025cbi, Sharma:2017eps, PhysRevD.111.083550, Cecchini:2023bqu},
or parity-violating terms like $f^2(\phi,R) F_{\mu\nu} \tilde{F}^{\mu\nu}$~\cite{Campanelli:2008kh,Jain:2012jy, Caprini:2014mja, Sharma:2018kgs, Bamba:2021wyx, Sharma:2018kgs}
Here, $R$ is the Ricci scalar, $F_{\mu\nu} = \partial_\mu A_\nu - \partial_\nu A_\mu$, and $A_\mu$ denotes the four-vector potential.

For our present purpose, we consider the simplest non-parity-violating magnetogenesis model. By appropriately choosing the conformal time ($\eta$) dependent coupling function $f(\phi,R) = f(\eta)$, we demonstrate that the model can induce large curvature fluctuations leading to primordial black hole (PBH) production, while simultaneously generating a large-scale magnetic field within the observable range. As mentioned earlier, we study scenarios where the produced PBHs survive until today and can act as either fully or partially viable dark matter candidates.

This paper is organized as follows. In \textbf{Sec.}~\ref{sec2}, we briefly describe the magnetogenesis model, discussing how it avoids both the strong coupling problem and the backreaction issue, while still generating the large-scale magnetic fields. We also provide a short discussion on the reheating dynamics, with particular emphasis on the perturbative reheating scenario. In \textbf{Sec.}~\ref{sec3}, we explain how the generated magnetic fields can induce curvature perturbations, and how the curvature spectrum depends on both the magnetogenesis parameters and the reheating dynamics. In \textbf{Sec.}~\ref{sec4}, we outline the PBH formation mechanism and analyze how the PBH mass and energy density fraction depend on the reheating dynamics as well as the magnetogenesis parameters. In \textbf{Sec.}~\ref{sec5}, we discuss the associated secondary gravitational waves that can arise from the anisotropic stress of the generated magnetic fields, which could be probed by future GW experiments. Finally, in \textbf{Sec.}~\ref{sec6}, we present our main results and conclusions, focusing on how a connection can be established between magnetogenesis scenarios and PBH formation, thereby opening a testable pathway linking primordial magnetic fields (PMFs) with scalar perturbations and cosmological observables.

\section{Model for Magnetogenesis}\label{sec2}
Inflation offers a compelling framework for generating large-scale magnetic fields from quantum fluctuations~\cite{Guth:1980zm, Linde:1981mu, Albrecht:1982wi, Starobinsky:1980te, PhysRevD.50.7222}. However, the standard electromagnetic (EM) action is conformally invariant in a Friedmann-Lemaître-Robertson-Walker (FLRW) background, leading to a rapid dilution of magnetic fields during inflation, typically scaling as $B \propto 1/a^2$. To generate and sustain cosmologically relevant magnetic field strengths, this conformal invariance must be broken. A wide range of mechanisms has been proposed to achieve this, but many suffer from severe backreaction or strong coupling problems~\cite{Tripathy:2021sfb}.

Although several approaches have been proposed to address the backreaction and strong coupling problems while still generating magnetic fields consistent with present-day observations~\cite{Sharma:2018kgs, Kobayashi:2019uqs, Haque:2020bip, Maiti:2025rkn, Maiti:2025awl, Maiti:2025cbi}, in this work we focus on a specific magnetogenesis scenario that naturally produces a broken power-law magnetic power spectrum within the current observational bounds. In particular, we consider a class of models featuring a sawtooth-like coupling function, where magnetic field amplification occurs predominantly during the reheating epoch~\cite{Maiti:2025awl}. In these scenarios, the coupling function begins from unity during inflation, increases to a maximum value by the end of inflation, and subsequently decreases back to unity by the end of reheating, thereby restoring the conformal invariance of the electromagnetic (EM) field in the post-reheating Universe~\cite{Maiti:2025awl, Sharma:2017eps, Papanikolaou:2024cwr}.

The evolution of the EM field in this framework is described by the action
\begin{align}
    \mathcal{S}_{\rm EM} = \frac{1}{4} \int d^4x\, \sqrt{g}\, I^2(\eta)\, F_{\mu\nu} F^{\mu\nu},
\end{align}
where \( I(\eta) \) is a time-dependent coupling function that breaks the conformal invariance of the EM field during the early Universe. 

A convenient parametrization of the coupling function, following Refs.~\cite{Maiti:2025awl, Sharma:2017eps, Papanikolaou:2024cwr}, is
\begin{align}\label{eq:f_coupling}
    I(\eta)=
    \begin{cases}
        (a/a_i)^{n}, & a_i < a \leq a_e, \\[6pt]
        \left(\dfrac{a_e}{a_i}\right)^{n} \left(\dfrac{a_e}{a}\right)^{n\beta}, & a_e \leq a \leq a_{\rm re}, \\[6pt]
        1, & a \geq a_{\rm re},
    \end{cases}
\end{align}
where the parameter \( \beta \equiv \Nk / \Nre \). Here, \( \Nk= \ln(\ae/a_i) \) represents the number of inflationary e-folds remaining after the pivot scale \( \kpv \) exits the horizon, and \( \Nre = \ln(\are/\ae) \) corresponds to the number of e-folds during reheating. The scale factors \( a_i \), \( \ae \), and \(\are\) denote the scale factor at the horizon crossing of the pivot scale and end of inflation and the end of reheating, respectively.

This parametrization provides a continuous and physically motivated evolution of the coupling function across different cosmological epochs, ensuring a smooth transition from the magnetogenesis phase to the standard electromagnetic regime. 
During inflation, we assume a de Sitter background, and the post-inflationary reheating phase is governed by the constant equation of state $\wre$. The scale factor evolves as 
\begin{align}\label{eq:f_coupling}
    a(\eta)=
    \l\{ 
    \begin{matrix}
        -\frac{1}{H_{\rm e} \eta} & a_i<a \leq \ae,\\
        \ae \left(\frac{\HI}{\delta} \right)^{\delta}\left(\eta -\ee + \frac{\delta}{\HI} \right)^{\delta} & \ae \leq a \leq \are, \\
    \end{matrix}
    \r.
\end{align}

where $\delta={2}/{(1 + 3 \wre)}$ and $w$ is the average equation-of-state (EoS) parameter of the background during reheating. 
$\ee$ conformal time at the end of inflation. $\ere$ defines the end of reheating. 
Note that to ensure restoration of conformal invariance, we set $I(\eta \geq \ere) = 1$. Imposing this condition, we obtain $\alpha =n\beta \delta$. Where $(\ee,\ere)$ mark the time for the end of inflation and end of reheating accordingly.
    
\textbf{Production during inflation:} This class of magnetogenesis models generates non-helical magnetic fields, where both polarization modes are equally excited~\cite{Sharma:2017eps, Papanikolaou:2024cwr}. Working in the Coulomb gauge, $A_0 = 0$ and $\partial^i A_i = 0$, and considering 
the Fourier transformation 
\begin{align}
    A_i(\vx,\eta) = \int \frac{d^3\vk}{(2\pi)^3}\, \epsilon_i^\lambda(k) 
    A_k^\lambda(\eta)\, e^{i \vk \cdot \vx}.
\end{align}
The equation of motion for the modified gauge field mode function 
$\mAk^{\lambda}(\eta) \equiv I(\eta)\, A_k^\lambda(\eta)$ is given by
\begin{align}\label{eq:A_k}
    \mAk'' + \left(k^2 - \frac{I''(\eta)}{I(\eta)} \right) \mAk = 0.
\end{align}
In the above equation, we have suppressed the polarization index $\lambda$, as both 
polarization modes evolve identically. There are two distinct epochs during which significant amplification of the gauge field occurs. The first stage of amplification takes place during inflation. In a de Sitter inflationary background, the solution for the gauge field mode function is~\cite{Maiti:2025awl}
\begin{align}
    \mAk(\eta)=\sqrt{-\frac{\pi\eta}{4}}e^{i(n+1/2)\pi/2}\mathrm{H}^{(1)}_{n+1/2}(-k,\eta) ,
\end{align}
where $\mathrm{H}^{(1)}_{n+1/2}(x)$ denotes the Hankel function of the first kind. To obtain this solution, we impose the Bunch-Davies initial condition, which assumes that all modes are initially deep inside the horizon. In this limit, i.e., for $(-k\eta) \rightarrow \infty$, the solution approaches a plane-wave form, $
\mAk(-k\eta \gg 1) \simeq {e^{-ik\eta}}/{\sqrt{2k}}$.
As inflation ends, the relevant cosmological modes are stretched beyond the horizon, and their evolution becomes effectively frozen. At this stage, one can compute the spectra of the comoving magnetic and electric energy densities. These are given, respectively, by~\cite{Maiti:2025awl, Maiti:2025cbi}
\begin{subequations} \label{6a}
   \begin{align}
    \tmPb(k,\ee)&=\frac{\ke^4}{8\pi}\frac{2^{2n+1}\Gamma^2(n+1/2)}{\pi^2}\l(\frac{k}{\ke}\r)^{\nb}\\
    \label{6b}
    \tmPe(k,\ee)&=\frac{\ke^4}{8\pi}\frac{2^{2n-1}\Gamma^2(n-1/2)}{\pi^2}\l(\frac{k}{\ke}\r)^{\nb+2} 
\end{align} 
\end{subequations}
where $\tmPb=a^4\mPb(k,\eta)$ and $\tmPe=a^4\mPe(k,\eta)$ denote the comoving energy density of the magnetic field and electric field per logarithmic wavenumber respectively. Here $\nb = 4 - 2n$ denotes the magnetic spectral index on super-horizon scales, i.e., for $k < \kre$.

\textbf{Production during reheating:} On the other hand, due to the non-trivial time evolution of the coupling function during reheating, there is an additional production of the gauge field in this phase. To analyze this, we solve Eq.~\eqref{eq:A_k} during the reheating epoch, and further assume the electrical conductivity during this is negligible. Depending on the comoving wavenumber $k$, the behavior of the gauge field modes differs depending on whether they are super-horizon ($k <\kre$) or sub-horizon ($k >\kre$) at the end of reheating. 
The general solution of the gaue field during reheating era is ~\cite{Maiti:2025awl}
\begin{align}
    \mAk^{\mathrm{re}}(k,\eta>\ee) = \sqrt{k\tilde{\eta}} \left\{ d_1 J_{\alpha+1/2}(k\tilde{\eta}) + d_2 J_{-\alpha-1/2}(k\tilde{\eta}) \right\}.
\end{align}  
Here we define $\tilde{\eta}=\eta-\ee+\frac{\delta}{\HI}$ and \( J_{\nu} \) are Bessel functions, and \( d_1 \) and \( d_2 \) are integration constants. These constants are determined by enforcing the continuity of \( \mAk \) and its first derivative at the transition point \( \eta = \ee \).  

Applying these conditions yields  
\begin{subequations}
    \begin{align}
        d_1 &= \frac{\pi \xe^{1/2}}{2\cos(\pi\alpha)} \left[ \mAki(\xe) \tilde{\mJ}_{-\alpha-1/2}(\xe) + {\mAki}'(\xe) \mJ_{-\alpha-1/2}(\xe) \right], \\
        d_2 &= \frac{\pi \xe^{1/2}}{2\cos(\pi\alpha)} \left[ \mAki(\xe) \tilde{\mJ}_{\alpha+1/2}(\xe) - {\mAki}'(\xe) \mJ_{\alpha+1/2}(\xe) \right],
    \end{align}
\end{subequations}  
Here $\mAki(\xe)$ is the solution of the gauge field defined at the end of inflation. Where we define \( \xe = -k\ee = k/\ke \), and the auxiliary functions are  
\begin{subequations}
   \begin{align}
    \tilde{\mJ}_{\alpha+1/2}(\xe) &= \frac{(1+\alpha)\mJ_{\alpha+1/2}(\xe) - \xe \mJ_{\alpha+3/2}(\xe)}{\xe}, \\
    \tilde{\mJ}_{-\alpha-1/2}(\xe) &= \frac{\alpha\mJ_{-\alpha-1/2}(\xe) + \xe \mJ_{-\alpha+1/2}(\xe)}{\xe}.
   \end{align}
\end{subequations}  
Since we are primarily interested in modes that exited the horizon well before the end of inflation, i.e., \( k < \ke \), we focus on the regime where \( \xe \ll 1 \).

As a result, the comoving electric and magnetic energy spectra exhibit different scaling behaviors in these two regimes. For the super-horizon modes ($k <\kre$), the spectra scale as~\cite{Maiti:2025awl},
\begin{subequations}
    \begin{align}
        \tmPb(k,\ere) &=\frac{\ke^4}{8\pi}\frac{\qnb^1(n)}{(\alpha+1/2)^2}\l(\frac{\ke}{\kre}\r)^{2(\alpha+1)}\l(\frac{k}{\ke}\r)^{\nb} ,\\
        \tmPe(k,\ere) &=\frac{\ke^4}{8\pi}\frac{\qnb^1(n)}{4}\l(\frac{\ke}{\kre}\r)^{2\alpha}\l(\frac{k}{\ke}\r)^{\nb-2} .
    \end{align}
\end{subequations}
In the above expression, we note that the electric field exhibits a decaying spectrum 
compared to the inflationary case, Eq.~\eqref{6b}, due to the decreasing value of the
coupling function during reheating. Whereas for the sub-horizon modes $(k>\kre)$, the magnetic and electric spectra behave as~\cite{Maiti:2025awl}
\begin{align}
    \tmPb(k>\kre,\ere)\simeq \frac{\ke^4}{4\pi}\frac{\qnb^2(n)}{\pi}\l(\frac{k}{\ke}\r)^{\nb-2(\alpha+1)} ,\\
    \tmPe(k>\kre,\ere)\simeq \frac{\ke^4}{4\pi}\qnb^2(n)\l(\frac{k}{\ke}\r)^{\nb-2(\alpha+1)} .
\end{align}
Where $\qnb^1(n)$ and $\qnb^2(n)$ is defined as
\begin{subequations}
    \begin{align}\label{eq:fnb1}
    \qnb^1(n)&=\frac{2^{2n+1}(\alpha-n)^2\Gamma^2(n+1/2)}{\cos^2(\pi\alpha)\Gamma^2(1/2-\alpha)\Gamma^2(\alpha+1/2)} ,\\
    \qnb^2(n)&=\frac{2^{2(n+\alpha)}(\alpha-n)^2\Gamma^2(n+1/2)}{\cos^2(\pi\alpha)\Gamma^2(1/2-\alpha)} .
\end{align}
\end{subequations}
After the end of reheating, the universe enters the standard radiation-dominated era, during which the electromagnetic (EM) field regains its gauge-invariant form, and no further amplification of magnetic fields occurs. In this phase, all fundamental particles have already been produced, and the universe exhibits extremely high electrical conductivity. In the high-conductivity limit, the rapid response of charged particles leads to the suppression of electric fields, while the magnetic fields evolve adiabatically. This implies that the comoving magnetic energy density remains constant with time, an assumption valid on sufficiently large scales where magnetohydrodynamic (MHD) effects are negligible.

In this scenario, therefore, the magnetogenesis model produces a broken power-law type magnetic spectrum, with spectral breaks appearing near $k\simeq\kre$, corresponding to the modes re-entering the Hubble horizon at the onset of the radiation-dominated era. During this time, for modes that are close to or larger than the size of the Hubble radius, the MHD evolution can be safely ignored. Under this assumption, the magnetic power spectrum at the beginning of the radiation-dominated era $(\eta=\ere)$ can be written as,
\begin{align}\label{eq:mPb_re}
    \mPb(k,\eta\simeq\ere)= \mB_{\rm p}^2\times\l\{
    \begin{matrix}
       \l(\frac{k}{\kre} \r)^{\nb} & \, k\leq\kre\\
        \l(\frac{k}{\kre} \r)^{\nb-2(\alpha+1)} & k>\kre,
    \end{matrix}
    \r.
\end{align}
where $\mB_{\rm p}$ is the amplitude of the magnetic field at peak during this time, and it can be expressed as
\begin{align}
    \mB^2_{\rm p}=\frac{\HI^4}{8\pi}q^1_{\nb}(n)\l(\frac{\ke}{\kre}\r)^{2(\alpha+1)-\nb}\l(\frac{\ae}{\are}\r)^4
\end{align}
Here $\ae$ and $\are$ are the scale factors defined at the end of inflation and the end of reheating, respectively.
As the peak of the magnetic field is located at $k\simeq\kre$, it is sensitive to the details of the reheating dynamics, which mainly depend on the reheating temperature $(\Tre)$. In our analysis, we work within a parameter space where the produced magnetic field energy density is always smaller than the background energy density, i.e., $\rhoem(\ere) < \rho_r(\ere)$,
where $\rho_r(\ere)$ denotes the background radiation energy density at the end of reheating.
\begin{figure*}[htbp!]
    \centering
    \includegraphics[width=0.47\linewidth]{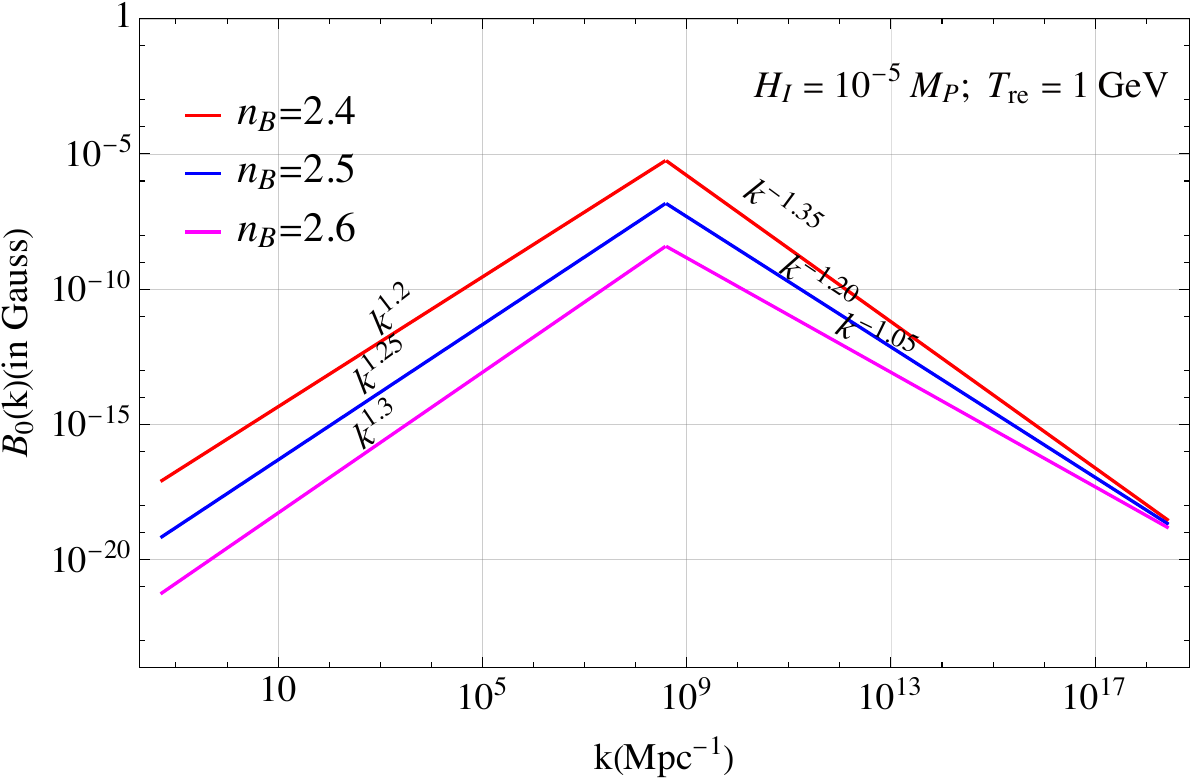}
    \includegraphics[width=0.47\linewidth]{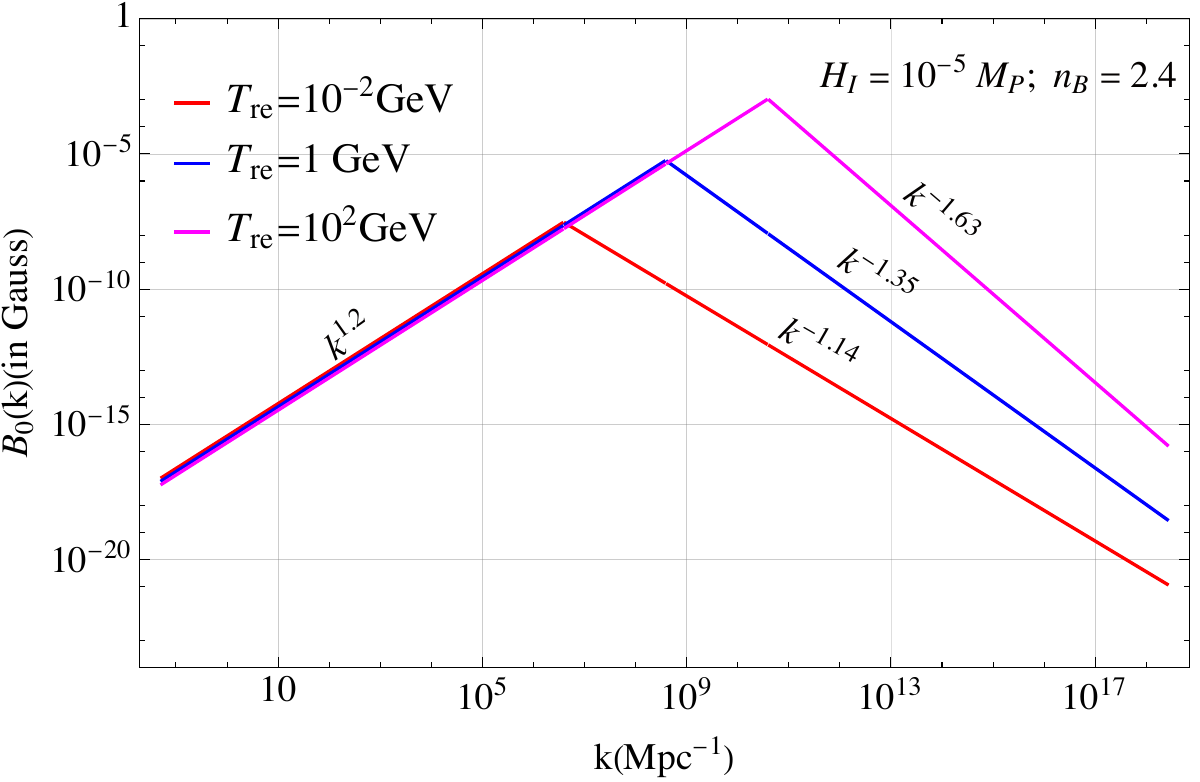}
    \caption{In this figure, we present the present-day magnetic field strength $B_0(k)$ (in Gauss) as a function of the comoving wavenumber $k$ (in $\Mpc^{-1}$). The left panel illustrates the dependence of the magnetic field spectrum on the magnetogenesis parameter $\nb = 4 - 2n$, with the reheating temperature fixed at $\Tre = 1\,\GeV$. In contrast, the right panel shows how the duration of the reheating era influences both the spectral shape and the amplitude of the magnetic field, for a fixed magnetic spectral index $\nb = 2.4$. In both cases, we assume an average equation of state during reheating of $\wre=0$ and the inflationary energy scale as $\HI\simeq 10^{-5}\Mp$. }
    \label{fig:mag}
\end{figure*}

\textbf{Relating inflationary and Reheating Parameters:} 
Inflation is a phase of accelerated expansion in which the Universe undergoes nearly exponential growth. During this period, the primordial seed fluctuations namely scalar, vector and tensor responsible for the large-scale structure, primordial magnetic field, gravitational waves of the Universe are generated when the corresponding modes cross the horizon. The duration of inflation depends on both the inflationary energy scale and the shape of the inflaton potential. 

Rather than focusing on a specific inflationary model, we consider a broad class of scenarios in which the duration of the inflationary era can be expressed in terms of the average equation of state (EoS) $\wre$ during reheating and the reheating temperature $\Tre$. In this work, we assume that reheating proceeds perturbatively~\cite{PhysRevLett.113.041302}. 

If we assume that all tensor fluctuations at CMB scales arise solely from primary gravitational waves (PGWs), then the inflationary Hubble parameter $\HI$ can be related to the tensor-to-scalar ratio $r$ and the amplitude of the scalar curvature power spectrum $\As$ via $\HI = \pi \Mp \sqrt{{r\,\As}/{2}}$.
Throughout our analysis, we take $r_{k_* = 0.05} \simeq 0.036$~\cite{Planck:2018jri, BICEP:2021xfz, BICEP2:2018kqh, Clarke:2020bil} and $\As \simeq 2.1 \times 10^{-9}$. Here $\Mp = M_{\rm pl}/\sqrt{8\pi} \simeq 2.4 \times 10^{18}~\GeV$ is the reduced Planck mass. 

If $\Nk$ is the $e$-folding number associated with the CMB pivot scale $k_*$, the highest comoving wavenumber that leaves the horizon at the end of inflation is $\ke = \kpv\, e^{\Nk}$. It can further be expressed in terms of $\wre$ and $\Tre$ as~\cite{Chakraborty:2024rgl, Maiti:2025cbi}
\begin{align}\label{eq:ke}
    \ke = \left( \frac{43\,\gsre}{11} \right)^{1/3}
    \left( \frac{\pi^2 \gsre}{90} \right)^{\sigma}
    \frac{\HI^{1 - 2\sigma}\, \Tre^{4\sigma - 1} \, T_0}{\Mp^{2\sigma}},
\end{align}
where $\sigma={1}/{3(1 + \wre)}$, $T_0 \simeq 2.725~\mathrm{K}$ is the present-day CMB temperature, and $\gsre = 106.75$ is the effective number of relativistic degrees of freedom at the end of reheating.
\begin{table*}[t]
 \begin{tabular}{|c ||c| c| c |c |c |c |c|} 
\hline
 $\nb$ & $2.1$ & $2.2$ & $2.3$ & $2.4$ & $2.5$ & $2.6$ & $2.7$  \\
 \hline
 $\Tre~(\text{in}~\GeV)$ & $0.01$ & $0.15$ & $4.11$ & $10^2$ & $2.3\times 10^3$ & $4.8\times 10^4$ & $9.05\times 10^5$ \\
 \hline
 $B_0(k=1~\Mpc^{-1})(\text{in Gauss}~)$ & $2.5\times 10^{-10}$ & $2.9\times 10^{-12}$ & $2.8\times 10^{-14}$ & $2.1\times 10^{-16}$ & $1.3\times 10^{-18}$ & $6.4\times 10^{-21}$ & $2.7\times 10^{-23}$ \\
 \hline
 \end{tabular}
 \caption{ In the table above, we have presented the lower reheating temperature $\Tre$ (in GeV)  to avoid the backreaction due to the excessive production of the gauge field, and we also listed the corresponding present-day magnetic field for the $1~\Mpc$ scales. Here we consider the inflationary energy scale as $\HI\simeq 10^{-5}\Mp$ and we consider that the EoS during reheating period is $\wre=0$ (matter like evolution).}
 \label{tab:b0_backreaction}
\end{table*}

Assuming no significant entropy production after reheating, the comoving entropy density remains constant, $a^3(t)s = \mathrm{const}$, where $s$ is the entropy density. Using this, the smallest comoving wavenumber re-entering the Hubble radius at the end of reheating can be written as~\cite{Chakraborty:2024rgl, Maiti:2025cbi}
\begin{align}\label{eq:kre}
    \kre = 3.9 \times 10^6 \left( \frac{\Tre}{10^{-2}} \right) \, \Mpc^{-1}.
\end{align}
In the above equation we have utilized the well known Hubble evolution $H = \HI (\ae/a)^{3(1+\wre)/2}$.
From the above, it is clear that changing $\wre$ or $\Tre$ modifies both the duration of inflation and the length of the reheating era. In our analysis, we assume a matter-like reheating phase with $\wre = 0$.

\textbf{Backreaction Issue:} 
Though out the computation we treated the gauge field as a perturbative component. Hence the total EM energy density is assumed to remain smaller than the background energy density, i.e., $\rho = 3H^2\Mp^2$, during both the inflationary and post-inflationary eras. Particularly the fractional energy density of the EM field at the end of reheating  satisfies $\delta_{\rm em} = {\rho_{\rm em}(\ere)}/{\rho_{\rm bg}(\ere)} \leq 1$,
which ensures that the produced EM field energy density always stays below the background energy density, even after reheating. 
Assuming the electric field decays almost instantaneously due to the high electrical conductivity of the Universe after the end of reheating, while the magnetic field evolves adiabatically, $\rho_{\rm B} \propto a^{-4}$, the $\delta_{\rm em}$ at the end of reheating assumes, 
\begin{align} \label{deltaless1}
    \delta_{\rm em}(\ere) =\frac{\HI^2}{24\pi\Mp^2}\frac{\qnb^1}{2(\alpha+1)-\nb}\l(\frac{\are}{\ae}\r)^{\frac{2\alpha-\nb}{2}}
\end{align}
To arrive at the above quantity, we have considered EoS during reheating is $\wre=0$. In addition to the above, we need to take into account the additional constraint on reheating temperature $\Tre > T_{\rm BBN}$, the Big-Bang Nucleosynthesis temperature. We further need to take into account the maximum possible bound on the magnetic field strength at $1$ Mpc scale being $\lesssim 1$ nG. To this end let us point out that in our defined fractional energy density at the end of reheating, we neglected MHD effects, since they are only relevant for modes that remain deep inside the horizon at that epoch~\cite{Brandenburg:2021bfx, Brandenburg:2021pdv, Sharma:2022ysf}. Therefore, the total energy density of the produced EM field is not affected by MHD.

\textbf{Present-day magnetic field and parameter constraints:} 
After reheating, the conformal symmetry of the gauge field is restored, and no further magnetic field production occurs. Once reheating is completed
the universe becomes radiation-dominated with large conductivity 
Owing to this high conductivity, the electric field component decays almost instantaneously, while the large-scale modes that remain outside the horizon evolve adiabatically. Therefore, the present-day magnetic field for different scales can be followed from the following  relation $a_0^2 B_0 = \are^2 B(\ere)$, and the eq.\ref{eq:mPb_re}
\begin{align}
    B_0(k)= \sqrt{\frac{\ke^4\qnb^1}{8\pi}}\l(\frac{\ke}{\kre}\r)^{\alpha+1}\l(\frac{k}{\ke}\r)^{\nb/2}.\label{B0now}
\end{align} 
We set all the scales $k$ with respect to this. Given the equation of state $\wre =0$ which is the case we will discuss through out the paper, the above constraining equation \ref{deltaless1} we obtain minimum possible reheating temperature tabulated in Tab.\ref{tab:b0_backreaction} for a given $\nb$ and $\HI = 10^{-5} \Mp$, and the corresponding present-day magnetic field strength at the comoving scale of $1$ Mpc.
For example, $\nb = 2.1$, we obtain that the minimum reheating temperature would be $\Tre\simeq0.01$ GeV, consistent with non-backreaction, and the present-day magnetic field turns out to be $\sim 10^{-10}$ G, consistent with the observational upper bound. 
 
Eq.\ref{B0now} clearly suggests that as $k/\kre <1$ for large scale magnetic field, increasing $\nb$ reduces the present day value of the magnetic field as $B_0(k) \propto (k/\ke)^{\nb/2}$. Therefore, to remain consistent with current observational bounds, the magnetic spectral index should be $\nb > 2.1 $. Further, if we consider the $\nb>2.5$, large-scale magnetic fields can still be generated, but their strength will fall below the present-day lower bound $B_0 \gtrapprox 10^{-19}$ G.

In Fig.~\ref{fig:mag}, we present the present-day magnetic field strength $B_0(k)$ (in Gauss) as a function of the comoving wavenumber $k$ (in $\Mpc^{-1}$). 
The left panel illustrates how $B_0(k)$ depends on the magnetogenesis parameter $\nb$, recalling that the magnetic spectral index is given by $\nb = 4 - 2n$. For this plot, we fix the reheating temperature to $\Tre = 1\,\GeV$, which results in a spectral break in the magnetic spectrum at $k \simeq 7.8 \times 10^8\,\Mpc^{-1}$. The three colored curves correspond to $\nb = 2.4$ (red), $\nb = 2.5$ (blue), and $\nb = 2.6$ (magenta).  
For $k < k_{\rm re}$ the magnetic spectrum scales as $B_0 \propto k^{\nb/2}$, while for $k > k_{\rm re}$ reheating effects modify the slope to $B_0 \propto k^{\nb/2-(\alpha+1)}$. For $\nb=2.4$, this gives a blue-tilted spectrum $B_0 \propto k^{1.2}$ on super-horizon scales and a red-tilted spectrum $B_0 \propto k^{-1.35}$ on sub-horizon scales, as shown in Fig.~\ref{fig:mag}. Increasing to $\nb=2.6$ shifts the tilts to $B_0 \propto k^{1.3}$ and $B_0 \propto k^{-1.05}$, respectively.


The right panel of Fig.~\ref{fig:mag} illustrates the impact of reheating duration on $B_0(k)$. For fixed $\nb$, the large-scale magnetic field amplitude remains unchanged, while the spectral peak shifts to higher $k$ as the reheating temperature increases. This shift originates from the spectral break at $k\simeq k_{\rm re}$: a higher $\Tre$ implies a shorter reheating duration, pushing $k_{\rm re}$ toward larger frequencies. Consequently, super-horizon modes preserve their scaling, whereas sub-horizon modes exhibit a $\Tre$-dependent tilt. For instance, at $\Tre=10^{-2}\,\GeV$ the sub-horizon spectrum follows $B_0\propto k^{-1.63}$, while at $\Tre=10^{2}\,\GeV$ it softens to $B_0\propto k^{-1.14}$. In both panels we assume $\wre=0$ and $\HI\simeq 10^{-5}\Mp$.

\section{Induced Curvature Power Spectrum}\label{sec3}
Primordial magnetic fields (PMFs), if generated in the early universe during inflation or the subsequent reheating phase, not only provide a viable origin for the large-scale magnetic fields observed today, but also act as sources of secondary gravitational waves~\cite{Sorbo:2011rz, Caprini:2014mja, Ito:2016fqp, Sharma:2019jtb, Okano:2020uyr, Maiti:2025cbi, Maiti:2025rkn, Maiti:2024nhv, Kumar:2025jfi} and curvature perturbations~\cite{Fujita:2013qxa, PhysRevD.94.043523}. In this class of magnetogenesis scenarios considered here, the magnetic field reaches its maximum strength at the end of reheating. Consequently,
we should expect that the corresponding curvature perturbation induced by the magnetic field will also peak shortly after reheating. The comoving curvature perturbation on a uniform density hypersurface is defined as $\zeta=-\Phi-\mH\frac{\delta\rho}{\rho'}$~\cite{Bassett:2005xm, Wands:2000dp, PhysRevD.94.043523, Baumann:2009ds}, where ${\cal H} = a'/a$, and prime is with respect to conformal time. Where, $\rho$ and $\delta \rho$ are the background energy density and its fluctuation respectively. There are two ways the produced magnetic field can source the curvature: it can directly contribute from the energy density fluctuations denoted as $\delta\rho=\delta\rhob$~\cite{Fujita:2013qxa, PhysRevD.94.043523}, and it can also source the curvature through the anisotropies in the energy-momentum tensor due to the magnetic field via the Bardeen potential $\Phi$~\cite{PhysRevD.81.043517, Saga_2020}. To compute the induced curvature power spectrum due to the magnetic field, we have broken it into two components, one is $\zeta_{\delta\rhob}=-\mH\frac{\delta\rhob}{\rho'}$ and the other is $\zeta_\Phi=-\Phi$. 
The associated curvature power spectrum can be written as  
\begin{equation}
\mPc(k)= \frac{k^3}{2\pi^2} | \zeta(k)|^2 
= \mPcb(k)+\mPcphi(k)+\mathcal{P}_{\zeta}^{\Phi\delta\rhoem}(k),
\end{equation}
Here $\mPcb$ indicates the contribution came from the density fluctuation of the magnetic field directly 
and $\mPcphi$ denotes the contributions that came from the anisotropic components of the energy-momentum 
tensor due to the magnetic field. At the scale of our interest $\mPcb \ll \mPcphi$, the contribution from the cross term $\mathcal{P}_{\zeta}^{\Phi\delta\rhoem}(k)$ correlating $\Phi$ and $\mH\frac{\delta\rhoem}{\rho'}$ will be negligibly small, and we ignore this term ono the term for simplicity.

To compute the curvature spectrum analytically, we utilize the fact that the maximum contribution will come from around the peak of the magnetic power spectrum at $k=\kre$, which is near the end of reheating. Therefore, around this scale we define the induced curvature perturbations $\zetab=-\delta\rhob/4\rho_{\rm ra}$, 
where we have used $\rho'_{\rm ra}\approx -4\mH \rho_{\rm ra}$ with $\rho_{\rm ra}$ being radiation energy density. Interesting to note that as both the magnetic field and radiation energy density dilute as $a^{-4}$ due to background expansion, $\zetab$ remains constant during the radiation-dominated era. The energy density fluctuations of the magnetic field in Fourier space as~\cite{Fujita:2013qxa, PhysRevD.94.043523}
\begin{align}\label{eq:delta_rhob}
    \delta\rhob(k,\eta)=\int \frac{d^3\vq}{(2\pi)^3}B(q,\eta)\cdot B(|\vk-\vq|,\eta) .
\end{align}
\begin{figure*}[htbp!]
    \centering
    \includegraphics[width=0.47\linewidth]{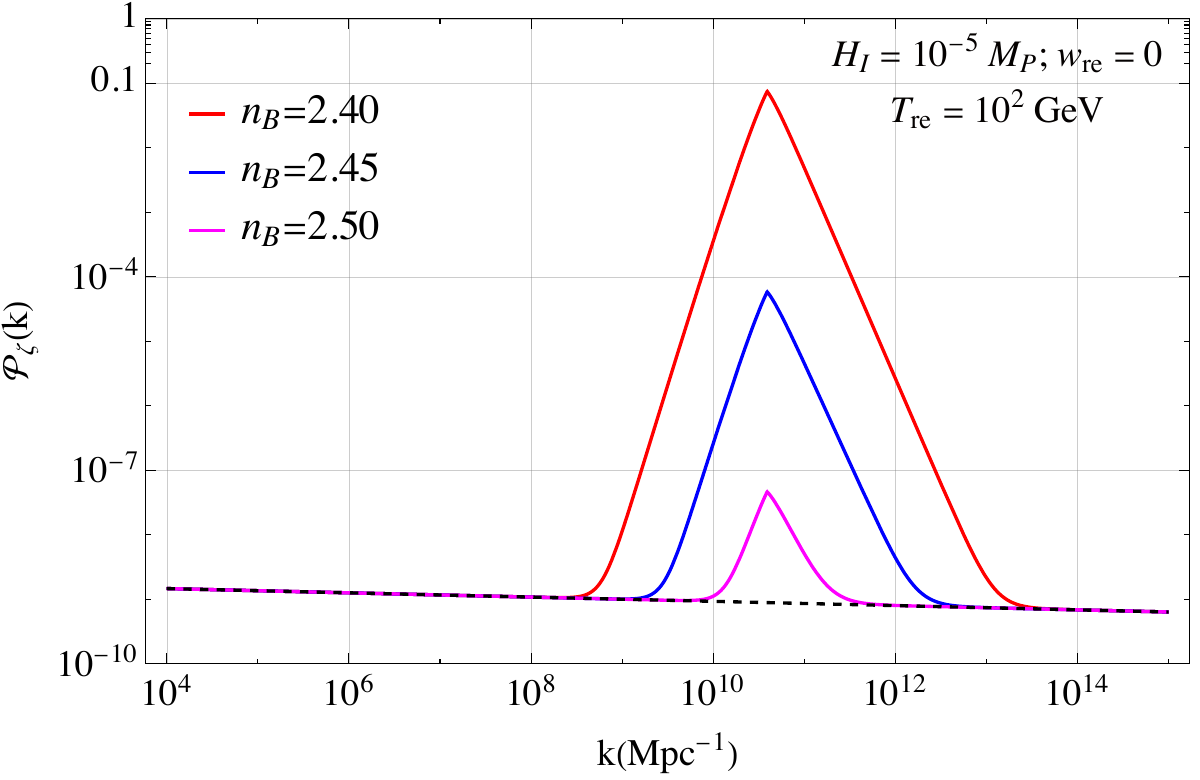}
    \includegraphics[width=0.47\linewidth]{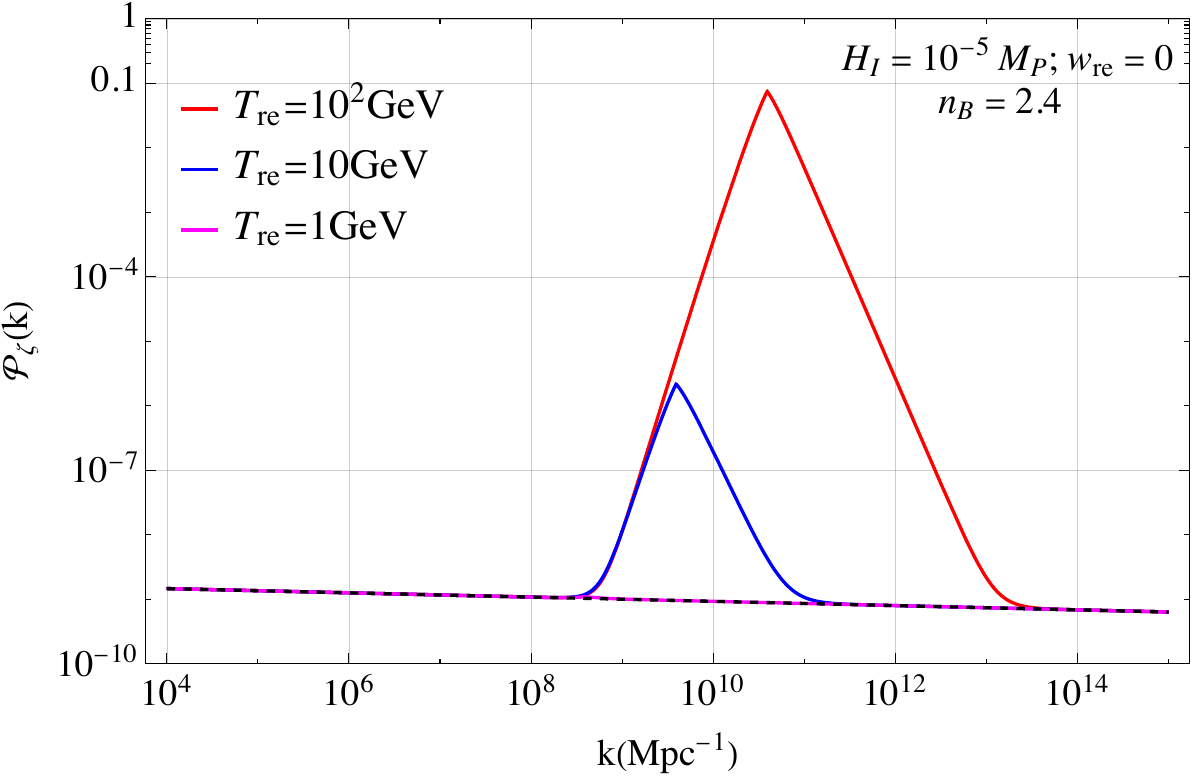}
    \caption{Here, we present the comoving curvature power spectrum $\mPc(k)$ as a function of the comoving wavenumber $k \, (\Mpc^{-1})$. 
In the left panel, we show the dependence on the magnetic spectral index $\nb$ for a fixed reheating scenario, specified by $(\wre = 0, \Tre = 10^2\,\GeV)$. 
In the right panel, we illustrate the impact of the reheating duration on $\mPc(k)$, where the EoS is fixed at $\wre = 0$ and the magnetic spectral index is set to $\nb = 2.4$. 
In both panels, the black dashed line represents the primary contribution to the curvature power spectrum due to the slow-roll inflationary dynamics.}
    \label{fig:pzeta}
\end{figure*} 
Utilizing this, we can define a two-point correlation function of the magnetic energy density fluctuation as
\begin{align}\label{eq:P_deltarhob}
    \langle \delta\rhob(\vk)\delta\rhob^{*}(\vk')\rangle=(2\pi)^3\delta^{(3)}(\vk-\vk')\frac{2\pi^2}{k^3}P_{\delta\rhob}(k).
\end{align}
Here $P_{\delta\rhob}(k)$ is the known power spectrum associated with the density fluctuation of the magnetic field. Combining Eq.\eqref{eq:delta_rhob} with Eq.\eqref{eq:P_deltarhob}, we get~\cite{Fujita:2013qxa, PhysRevD.94.043523}
\begin{align}\label{eq:P_deltab}
    P_{\delta\rhob}(k)=\int \frac{\d q}{q}\int_{-1}^1\d \gamma\,\frac{(1-\gamma^2)\mPb(q)\mPb(\vk-\vq|)}{\l(1+\frac{q^2}{k^2}-2\gamma\frac{q}{k}\r)^{5/2}},
\end{align}
where $\gamma=\hat{k}\cdot\hat{q}=\cos(\theta)$ where $\theta$ is the angle between two vector $\vk$ and $\vq$. Here we defined $u=q/k$.
Now, the induced curvature power spectrum associated with the magnetic energy density fluctuation defined at the end of reheating is
\begin{align}\label{eq:P_zeta_deltab}
    \mPcb(k)=\frac{1}{16}\frac{P_{\delta\rhob}(k)}{\rho_{\rm ra}^2} .
\end{align}
To compute the exact nature of the curvature power spectrum, we have done numerical integration. However, the amplitude of the spectrum at peak frequency can be analytically estimated as 
\begin{align}\label{eq:curv_peak}
  A_\zeta \simeq \frac{1}{24}\l(\frac{\qnb^1}{24\pi}\r)^2\l(\frac{\HI}{\Mp}\r)^4\l(\frac{\are}{\ae}\r)^{2\alpha-\nb}.
    \end{align}
Recall that $\delta_{\rm em}(\ere)$ is the fractional energy density of the magnetic field defined at the end of reheating. 

Apart from this we have additional secondary contribution from the modified dynamics of Bardeen potential $\Phi$ sourced by the anisotropic component of the magnetic field~\cite{PhysRevD.88.083515, PhysRevD.81.043517, Saga_2020}. After the reheating, our standard radiation-dominated era begins, and the peak of the frequency of the magnetic field is now inside the horizon. To track the evolution of the curvature perturbation due to magnetic field anisotropy, we have to solve the Bardeen potential equation in the presence of a source. For simplicity, we ignored the contribution originating from the entropy perturbation due to the magnetic field. In the radiation-dominated era, the evolution of the Bardeen potential sourced by the magnetic anisotropies is simply governed by~\cite{PhysRevD.88.083515, PhysRevD.81.043517, Saga_2020}
\begin{align}\label{eq:Phi_evo}
    \Phi''+\frac{4}{\eta}\Phi'+\frac{k^2}{3}\Phi=-\frac{\mH^2}{k^2}\l[ \frac{k^2}{3} -2\mH' \r]\Pi_{\rm B}.
\end{align}
In the above expression, we made use of ${\cal H} = 1/\eta$ during the radiation-dominated phase. Further, $\Pi_{\rm B}(k,\eta)$ is the source term associated with the magnetic field,
where $|\Pi_{\rm B}(k)|^2=\langle\Pi_{\rm B}(k)\Pi^{*}_{\rm B}(k)\rangle$ is define as~\cite{PhysRevD.81.043517, Saga_2020} 
\begin{align}
    |\Pi_{\rm B}(k)|^2=\frac{36\pi^3}{16\rho^2 a^8}\int\frac{\d q}{q}\frac{\mPb(q)\mPb(|\vk-\vq|)}{|\vk-\vq|^3} \,\mF(k,q,\gamma)
\end{align}
Here $\mPb(k)$ is the magnetic power spectrum defined in Eq.\eqref{eq:mPb_re}. Here $\mF(k,q,\gamma)$ defined as~\cite{PhysRevD.81.043517}
\begin{align}
    \mF(k,q,\gamma)=\frac{1+\gamma^2+u(2\gamma-6\gamma^3)+u^2(5-12\gamma^2+9\gamma^4)}{4(1+u^2-2u\gamma)}
\end{align}
To solve the above Eq.\eqref{eq:Phi_evo}, we have used the Green's function method, where the solution of the Bardeen potential $\Phi$ in the radiation-dominated era can be written as
\begin{align}
   \Phi(k,\eta_\nu)=\int_{\ere}^{\eta_\nu} \d\eta_1 \,\l(\frac{\mGk(\eta_\nu,\eta_1)}{\eta_1^2}\r)\times\l[ \frac{1}{3} -2\frac{\mH'}{k^2}  \r]\Pi_{\rm B} .
\end{align}
Here $\eta_\nu$ is the conformal time when the neutrinos are decoupled from the background thermal bath. Recall the fact that as the neutrino's anisotropy balances the magnetic anisotropy, the contribution from the magnetic anisotropy after neutrino decoupling is negligible~\cite {PhysRevD.70.043011, KAGRA:2021kbb}. Here, the Green's function associated with the differential operator corresponding to the left hand side of the eq.\ref{eq:Phi_evo} is
\begin{align}
    \mGk(\eta,\eta_1)&=\Theta(\eta-\eta_1)\frac{\pi\sqrt{3}\eta_1}{2\sin(3\pi/2)}\l(\frac{\eta_1}{\eta}\r)^{3/2}\nn\\
    &\times\l[\bJ_{3/2}(\tx)\bJ_{-3/2}(\tx_1)-\bJ_{3/2}(\tx_1)\bJ_{-3/2}(\tx) \r] .
\end{align}
In the above we define a new variable $\tx = \sqrt{3}\,k\eta$. In this magnetogenesis scenario, a magnetic field is produced starting from inflation to the entire period of reheating era, and post-reheating gague field production ceases to exist. Owing to the high electrical conductivity, the electric field decays immediately after the reheating ends, while the magnetic component evolves adiabatically. During radiation domination, both the background and the magnetic field dilute as $a^{-4}$ due to the background expansion, so the quantity $\Pi_{\rm B}(k,\eta)$ remains constant in time. The amplitude of the comoving magnetic field is constant on super-horizon scales, and it oscillates with constant amplitude for sub-horizon modes (recall that MHD affects modes deep inside the horizon, whereas our interest is primarily at the beginning of the radiation-dominated era~\cite{Sharma:2022ysf, Brandenburg:2021pdv}).
As the peak of wave number are inside the horizon during the radiation dominated era beging as magnetic field get its peak at $k\simeq\kre$, we recall that $\kre$ the lowest-mode that can able to re-enter into the horizon at the end of reheating, and PBH is very sensitive and mainly depend on the peak of the curvature power spectrum so if we consider only the dominating term then we can write the power spectrum of the Bardeen potential define as $P_{\Phi}(k,\eta)=\frac{k^3}{2\pi^2}\Phi^2(k,\eta)$ as 
\begin{align}\label{eq:P_Phi}
    P_{\Phi}(k,\eta)=\frac{k^3}{2\pi^2}\l[-\frac{1}{3} \int_{\ere}^{\eta}\frac{\d \eta_1}{\eta_1^2}\mGk(\eta,\eta_1)\r]^2 |\Pi_{\rm B}(k)|^2.
\end{align}
 Exact analytical computing of the curvature power spectrum due to the magnetic anisotropy is quite difficult. We have to do the above integral defined in Eq.\eqref{eq:P_Phi} numerically. At the beginning of the radiation-dominated era, when the peak amplitudes are just inside the horizon, the curvature power spectrum associated with the Bardeen potential is approximated as
 $\mPcphi(k,\eta\simeq \ere)\simeq \pi\mathcal{I}^2(\eta,\ere)\mPc^{\delta\rhob}(k)$, where $\mathcal{I}(\eta,\ere)$ is define as
\begin{align}
    \mathcal{I}(\eta,\ere)=\int_{\ere}^{\eta} \frac{\d \eta_1}{\eta_1^2}\mGk(\eta,\ere),
\end{align}
where $\mPc^{\delta\rhob}(k)$ is the induced curvature power spectrum arising from the magnetic-field energy density fluctuations, as defined in Eq.~\eqref{eq:P_zeta_deltab}. 
For super-horizon scales, the time integral appearing in the source term can be approximated as 
$\mathcal{I}(\eta,\ere)\simeq \ln(\eta/\ere)$ 
in a radiation-dominated background. This logarithmic behavior is well known for magnetically induced curvature perturbations~\cite{PhysRevD.88.083515, PhysRevD.81.043517, Saga_2020}.

As mentioned earlier, we focus on the beginning of the radiation-dominated era, during which primordial black holes (PBHs) are formed. Since, at this epoch, the peak frequency lies inside the horizon, the peak amplitude of the curvature perturbations induced by the magnetic field must be computed numerically at the onset of radiation domination. For instance, choosing $\nb = 2.40$ with a reheating temperature $\Tre = 10^2~\GeV$, we obtain 
$\mPcb(k_{\rm peak}) \simeq 0.11$, 
while the inflationary contribution is 
$\mPcphi(k_{\rm peak}) \simeq 0.014$. 
On the other hand, for $\nb = 2.45$ with the same reheating temperature, the induced curvature power spectrum is significantly suppressed, yielding 
$\mPcb(k_{\rm peak}) \simeq 8.89 \times 10^{-5}$ 
and 
$\mPcphi(k_{\rm peak}) \simeq 1.08 \times 10^{-5}$.

The spectral behavior of the total curvature power spectrum can be approximated by a broken power-law form. Modes that are far outside the horizon at the end of reheating exhibit a slightly red-tilted spectrum, 
$\mPc(k \ll \kre)\propto k^{\ns-1}$, 
where the inflationary curvature perturbations dominate. Since the electromagnetic field spectrum peaks around $k\simeq\kre$, the curvature power spectrum also develops a peak near this scale. For modes that remain just outside the horizon at the end of reheating, the spectrum approximately behaves as 
$\mPc(k\leq\kre)\propto k^{2\nb}$. 
In contrast, modes that re-enter the horizon around the end of reheating, i.e.\ $k\geq\kre$, scale as 
$\mPc(k\geq\kre)\propto k^{2\nb-4(\alpha+1)}$. 
For modes deep inside the horizon, the magnetic-field contribution becomes subdominant due to the decay of the magnetic energy density, and the spectrum again approaches the inflationary form, 
$\mPc(k \gg \kre)\propto k^{\ns-1}$.

Combining these behaviors, the curvature power spectrum can be approximated as
\begin{align}
    \mPc(k)\simeq 
    \l\{
    \begin{matrix}
        \As (k/k_{*})^{\ns-1} & k_{*}\leq k<\ksb,\\
        A_\zeta (k/\kre)^{\nc} & \ksb <k\leq\kre,\\
        A_\zeta (k/\kre)^{\nb-2(\alpha+1)} & \kre<k<\ksbs,\\
        \As (k/k_{*})^{\ns-1} &\ksbs<k<\ke,
    \end{matrix}
    \r.
\end{align}
Where $\nc=2\nb$ for $\nb\leq 3/2$, whereas for very stiff values of the magnetic spectral index, i.e., $\nb>3/2$, it takes $\nc=3.0$.
Here $\As\simeq 2.1\times 10^{-9}$ denotes the amplitude of the curvature power spectrum at CMB scales~\cite{Planck:2018jri}, and $A_\zeta$ is defined in Eq.~\eqref{eq:curv_peak}.  $\ksb$ and $\ksbs$ correspond to the mode associated with the first and second Spectral Breaks (SB), respectively, in the above spectrum. For $k>\ksb$, the secondary (magnetically induced) contribution dominates, while for $k>\ksbs$ the inflationary curvature power spectrum again becomes the dominant component.

In Fig.~\ref{fig:pzeta}, we present the comoving induced curvature power spectrum as a function of the comoving wavenumber $k$ for two distinct scenarios. In the left panel, we fix the reheating dynamics by choosing $\wre = 0$ and $\Tre = 10^2\,\GeV$. The three different colors correspond to $\nb = 2.40, 2.45$, and $2.50$. We find that the peak amplitude changes significantly with $\nb$ since the magnetic field strength is highly sensitive to the magnetic spectral index. In particular, for fixed reheating scenarios, there exists a threshold value of $\nb$ below which the peak amplitude of the induced curvature spectrum can reach $\mPc(k_{\rm peak}) \geq 10^{-2}$, providing the conditions necessary for PBH formation.

In the right panel of Fig.~\ref{fig:pzeta}, we show the effect of the reheating duration on the induced curvature power spectrum for a fixed magnetic spectral index $\nb = 2.40$. The three colors lines correspond to reheating temperatures $\Tre = 1,\GeV$ (magenta), $10,\GeV$ (blue), and $10^2,\GeV$ (red). We observe that as $\Tre$ increases, the peak of the curvature power spectrum shifts toward higher-frequency modes and its amplitude grows. This behavior originates from our chosen coupling prescription, in which the coupling is restored to unity at the end of reheating. Under this condition, the effective post-inflationary coupling $\alpha$ takes larger values for higher $\Tre$, leading to more efficient gauge field production at the peak wavenumber in high-$\Tre$ scenarios compared to low-$\Tre$ ones.

\section{PBH formation}\label{sec4}
Primordial black holes (PBHs) are assumed to have formed in the early universe due to the collapse of overdense regions as they re-entered the Hubble horizon. Unlike black holes formed through stellar evolution, PBHs originate from the dynamics of the early universe itself. During the post-inflationary era, sufficiently large curvature perturbations can re-enter the horizon and collapse under their own gravity to form PBHs. 
The collapse is only efficient if the local density contrast, defined as \( \delta = \delta \rho / \rho \), exceeds a critical threshold \( \delta_c \). Numerical simulations indicate that in a radiation-dominated background, the threshold value is approximately \( \delta_c \sim 0.4 \)~\cite{Escriva:2019phb, PhysRevD.105.124055}. If the local perturbation satisfies \( \delta \geq \delta_c \), the region can undergo gravitational collapse to form a PBH.
In our scenario, we find that the sawtooth-type magnetogenesis model leads to an enhanced magnetic field energy density peaked at the comoving scale \( k \simeq \kre \). The associated curvature perturbation spectrum reaches a peak amplitude of \( \mathcal{P}_\zeta(k \simeq \kre) \simeq \mathcal{O}(0.1) \), which is sufficient to trigger PBH formation through gravitational collapse. 
The mass of the PBHs are typically equal to the horizon mass at the time of horizon re-entry of the corresponding mode, where the horizon mass at the formation time is expressed as $M_{H}=\frac{4\pi}{3}\rho(\th)H^{-3}(\th)$,
where $\th$ is the cosmic time when the PBHs are formed and $\rho(\th)$ is the total background energy density during the formation time. Now, if we consider the efficiency factor accounting for the collapse dynamics, then we can write the PBH mass as $\Mpbh\simeq 4\pi\gamma\Mp^2H^{-1}(\th)$.
We recall that $\Mp=1/\sqrt{8\pi G}$ is the reduced Planck mass. Here, $H(\th)$ is the Hubble constant during the formation of the PBHs.

The PBH abundance is exponentially sensitive to the peak amplitude of the induced curvature power spectrum, whereas the mass of the PBHs typically depends on the peak frequency of the curvature power spectrum. The fractional energy density of the produced PBHs at the time of formation is quantified as $\beta(\Mpbh)=\rhopbh(\th)/\rho(\th)$,
where $\rhopbh(\th)$ is the initial energy density of the PBHs at the time of formation. Thus, we can express $\beta(\Mpbh)$ as~\cite{Khlopov:1980mg, Bullock:1996at, Garcia-Bellido:1996mdl, Yokoyama:1998pt, Kawasaki:1997ju, Garcia-Bellido:2017mdw, PhysRevD.103.083510, Bhaumik:2020dor, Solbi:2021wbo, Figueroa:2021zah, Frolovsky:2022qpg, Motohashi:2017kbs, Byrnes:2018txb,Cheng:2016qzb,Cheng:2018yyr, Ballesteros:2018wlw, Raveendran:2022dtb, Ragavendra:2020sop, 
Braglia:2020eai, Karam:2022nym}
\begin{align} \label{betap}
   \beta(\Mpbh)=\int_{\delta_c}^{\infty}\d \delta\,P(\delta) .
\end{align}
Here, $P(\delta)$ is the probability distribution of the density contrast $\delta$.

Here $P(\delta)$ is the probability distribution of the density contrast originated from the magnetic field. We recall that the density contrast is defined as $\delta=\delta\rhob/\rho_{\rm rad}\propto B^2$, so the $\delta$ is related to the energy density of the electromagnetic field, which depends quadratically on the underlying Gaussian magnetic field and electric field. So we can say that when the density contrast associated with the electromagnetic field is intrinsically non-Gaussian in nature~\cite{Saga_2020}. To capture the full non-Gaussian nature of the distribution function, one can use the MCMC analysis as discussed in \cite{Saga:2020ics}, but in our computation, we consider a generalized hyperbolic (GH) distribution function, which can be parameterized as
\begin{align}
    P_{\rm NG}(x) = C\,\exp\!\left[-\frac{3}{2}\sqrt{1+\frac{x^2}{b^2}}\right],
\end{align}
where \(C\) and \(b\) are two free parameters, which are fixed by imposing
\begin{align}
    \int_{-\infty}^{\infty} P_{\rm NG}(x)\,\d x = 1,
    \qquad
    \int_{-\infty}^{\infty} x^2 P_{\rm NG}(x)\,\d x = \sigma_B^2.
\end{align}
Using these relations, one finds \(C = 2.13\sigma_B^{-1}\) and \(b = 0.84\,\sigma_B\)~\cite{Kushwaha:2024zhd}, 
where \(\sigma_B\) is the standard deviation of the density fluctuations on the 
scale corresponding to the mass \(M\). The variance of the smoothed density contrast 
on the scale \(R = 1/k\) is

\begin{align}    \sigma^2_B=\frac{16}{81}\int \frac{\d q}{q}(qR)^4W^2(qR){\cal P}_{\zeta}(q) ,
\end{align}
\begin{center}
\begin{table*}[t]
\centering
    \begin{tabular}{|c ||c |c| c| c| c| c| }
    \hline
    Benchmark Point(BP) & BP1& BP2& BP3& BP4& BP3 & BP6\\
    \hline
    $\Tre(~\text{in}~\GeV)$ & $1$ & $1$ & $10^3$ & $10^3$ & $10^5$ & $10^5$\\
    \hline
    $\nb$ & $2.28$ & $2.285$ &  $2.5$ & $2.505$ & $2.655$ & $2.660$ \\
    \hline
    $B_0(1~\Mpc^{-1})(\text{in}~G)$ & $7.3\times 10^{-14}$ & $5.83\times 10^{-14}$ & $ 1.40\times 10^{-18}$ & $1.09\times 10^{-18}$ & $3.67\times 10^{-22}$ & $ 2.85\times 10^{-22}$ \\
    \hline
    $\fpbh(k_{\rm peak})~(\delta_c=0.45)$ & $7.56\times 10^{-4}$& $5.9\times 10^{-10}$ &  $1.12\times 10^{-3}$ & $1.38\times 10^{-11}$ & $1.47\times 10^{-4}$ & $1.27\times 10^{-12}$ \\
    \hline
    \end{tabular}
         \caption{In the above table, we list representative benchmark points for PBH production, along with the corresponding present-day magnetic field strength at the $1~\mathrm{Mpc}$ scale and the maximum value of $\fpbh$ at its peak. To obtain this, we have consider $\wre=0$ and the inflationary energy scale $\HI\simeq 10^{-5}\Mp$.}
    \label{tab:param_est}
 \end{table*}
\end{center}
\begin{figure}[t]
    \centering
    \includegraphics[width=0.95\linewidth]{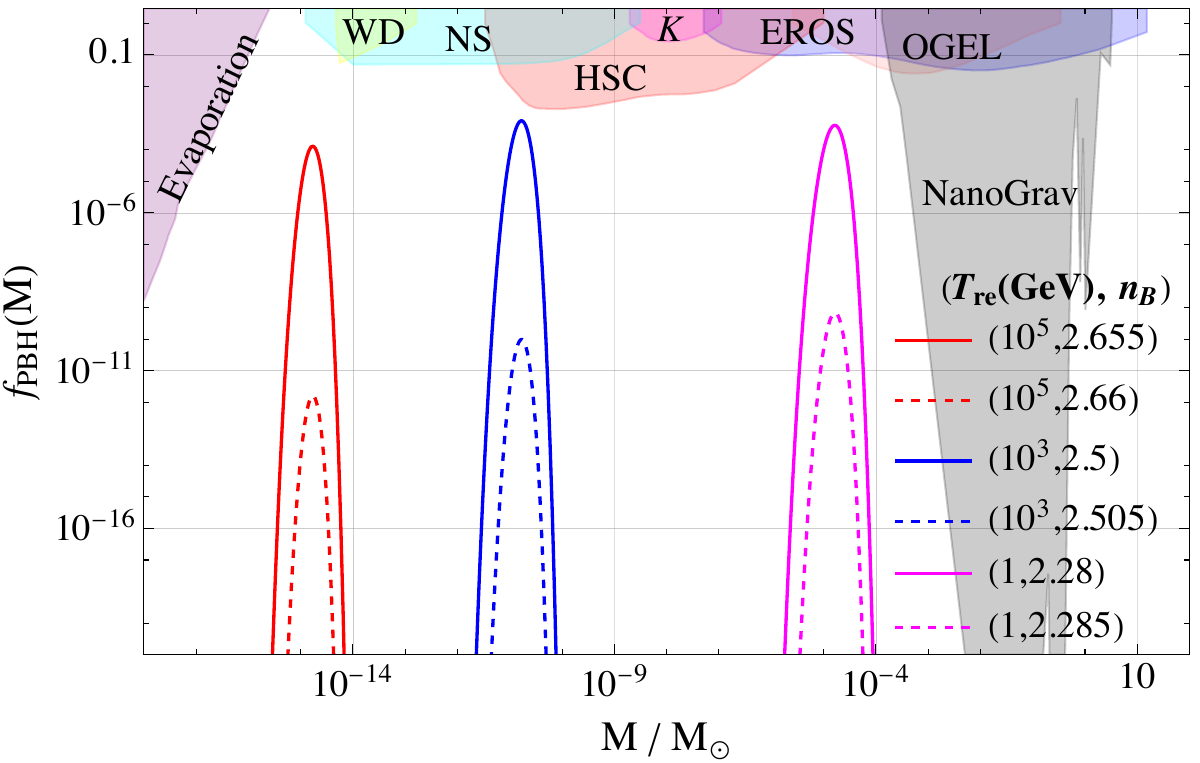}
    \caption{In the figure, we show the primordial black hole (PBH) dark matter fraction as a function of PBH mass, expressed in solar-mass units. We consider three fixed reheating temperatures, indicated by different colors. For each temperature, the solid and dashed curves correspond to two different choices of the magnetic spectral index. In all cases, we adopt a critical density contrast of $\delta_c = 0.45$. Table~\ref{tab:param_est} lists the corresponding present-day magnetic field strength at the $1\,\mathrm{Mpc}^{-1}$ scale, as well as the peak value of the PBH dark matter fraction.}
    \label{fig:fpbh}
\end{figure} 
\begin{figure}[t]
    \centering
    \includegraphics[width=0.95\linewidth]{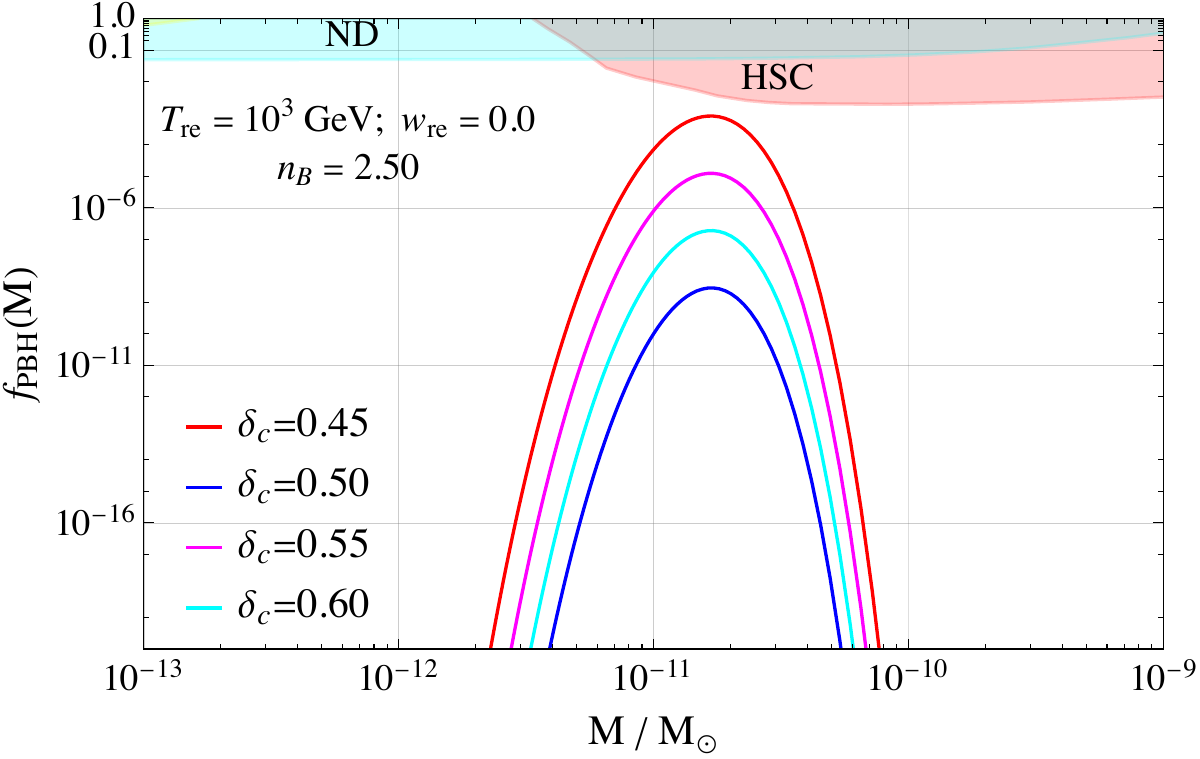}
    \caption{In the figure, we plot the primordial black hole (PBH) dark matter fraction as a function of PBH mass in solar-mass units. We illustrate the dependence of the PBH abundance on the threshold value of the density contrast, $\delta_c$, for a fixed reheating parameter with $\Tre = 10^3\,\GeV$ and $\wre = 0$. For all curves, we adopt a magnetic spectral index of $\nb = 2.50$.}
    \label{fig:fpbh_deltac}
\end{figure} 
where we commonly used window function $W(k'R)$ as a Gaussian, i.e., $W(k'R)=\exp(-k^2R^2/2)$ and utilizing this function we get
\begin{align}\label{eq:sigma}
    \sigma^2(\rm{M})=\frac{16}{81}\int_{0}^{\infty}\frac{\d q}{q}(qR)^4e^{-q^2R^2}\mPc(q) .
\end{align}
Once the behavior of the induced curvature perturbation $\mPc$ is determined, the initial
fractional energy density of PBHs at the time of formation can be computed using
Eq.~\eqref{betap}. We emphasize that the resulting PBH abundance is highly sensitive both to
the choice of the probability distribution function $P_{\rm NG}(x)$ and to the critical
density contrast $\delta_c$.

It is important to note that the presence of a magnetic field can modify the conventional
collapse criterion due to the additional magnetic pressure and its complex, nonlinear
interaction with the background plasma. Determining the exact value of the collapse
threshold in the presence of magnetic fields would require incorporating all relevant
contributions to the collapse equations and solving the full set of relativistic
magnetohydrodynamic equations, which is beyond the scope of the present work. Nevertheless,
an approximate estimate of $\delta_c$ in the presence of magnetic anisotropies has recently
been obtained in Ref.~\cite{Kushwaha:2024zhd}.

The argument proceeds as follows. Within an isotropic approximation, the total pressure and
energy density in the presence of a magnetic field are given by
$p_{tot}=p_{\rm rad}+p_B$ and $\rho_{\rm tot}=\rho_{\rm rad}+\rho_B$, respectively, where
$r_B=\rho_B/\rho_{\rm rad}$. The corresponding effective equation of state can then be written as
$w_{eff}=p_{\rm tot}/\rho_{\rm tot}=c_s^2+\alpha r_B$, where the coefficient $\alpha$ encodes
the dynamical evolution of the magnetic field on the relevant length scales. In
Ref.~\cite{Kushwaha:2024zhd}, the authors analytically estimated $\alpha \approx 2$, although
this value is expected to be sensitive to nonlinear magnetohydrodynamic effects.

In view of these theoretical uncertainties, we do not fix a single value of $\delta_c$.
Instead, we explore a representative range $\delta_c \in [0.45,\,0.60]$ in our subsequent
analysis and examine how variations in $\delta_c$ affect both the PBH mass spectrum and the
resulting abundance.

Now, once we know the behaviour of $\mPc(q)$ near the peak, Eq.~\eqref{eq:sigma} 
allows us to compute the associated variance of the distribution and thereby 
determine the initial abundance of PBHs. The mass of the PBHs formed at horizon 
entry can be expressed as a function of the comoving wavenumber $k$ as
 \begin{align}
     \Mpbh\simeq 1.57\times 10^{12}\,M_{\odot} \,\l(\frac{\gamma}{0.2}\r)\l(\frac{106.75}{g_{*,k}(T)}\r)^{1/6}\nn\\
     \times\l(\frac{g_{*,\rm eq}(T_0)}{3.81}\r)^{2/3}\l(\frac{1\Mpc^{-1}}{k}\r)^2 .
 \end{align}
 Here $g_{*,k}$ is the effective degree of freedom at the time of formation of the PBHs.
In our scenarios, the peak of the induced curvature power spectrum is situated at \( k \simeq \kre \). We recall that \( \kre \) is the lowest comoving wavenumber that is able to re-enter the horizon at the end of reheating, which mainly depends on the reheating temperature \( \Tre \) through Eq.~\eqref{eq:kre}. Utilizing the relation, $\kre = 3.9 \times 10^8 \left( {\Tre}/{1\,\GeV} \right) \, \Mpc^{-1}$,
if one assumes the reheating temperature being lower than \( \Tre < 10^7\,\GeV \), we can generate PBHs with masses even larger than \( \Mpbh > 10^{15}\,\mathrm{g} \), implying that such PBHs can survive until today and contribute as a cold dark matter (CDM) candidate.

We now define an observable quantity: the present-day mass fraction of dark matter in PBHs as $
\fpbh(\Mpbh) = {\Opbh(\Mpbh)}/{\Odm} $.
Here, \( \Odm \simeq 0.27 \) refers to the present-day energy density fraction of cold dark matter in the universe, normalized by the critical density.

After their formation, PBHs behave as non-relativistic matter, meaning their energy density redshifts as \( \rhopbh \propto a^{-3} \), whereas the background radiation energy density redshifts as \( a^{-4} \). As a result, the PBH energy density grows relative to radiation after formation. In our scenarios, since PBHs are produced during the radiation-dominated era, the present-day PBH abundance as a cold dark matter component can be expressed as~\cite{Raveendran:2022dtb, Ragavendra:2020sop}
\begin{align}
    \fpbh(\Mpbh)=\l(\frac{\gamma}{0.2}\r)^{3/2}\l(\frac{\beta(\Mpbh)}{1.46\times 10^{-8}}\r)\nn\\
    \times\l(\frac{g_{*,k}}{g_{*,\rm eq}}\r)^{-1/4}\l(\frac{\Mpbh}{M_{\odot}}\r)^{-1/2} .
\end{align}
Here $g_{*,k}\simeq 106.75$ and $g_{*,\rm eq}=3.36$ are the relativistic degrees of freedom at the time of PBH formation and matter-radiation equality.

In Tab.~\ref{tab:param_est}, we present a set of benchmark parameters used to compute the PBH fraction as a dark matter candidate predicted by this model. We consider three distinct values of the reheating temperature, $\Tre = 1,\,10^3,$ and $10^5~\mathrm{GeV}$, assuming a matter-like background during the reheating era. To study the dependence on the magnetic parameters, we adopt two different values of the magnetic spectral index, $\nb$, and examine their impact on the present-day PBH fraction.

For each parameter set, we also compute the corresponding present-day magnetic field strength at the $1\,\mathrm{Mpc}$ scale. Our results show that the PBH fraction is highly sensitive to the magnetic spectral index, while the present-day magnetic field varies more mildly. This strong sensitivity arises because the initial PBH abundance depends exponentially on the magnetic parameters (as seen in Gaussian-type PDFs); thus, even a small change in $\nb$ leads to a substantial variation in the resulting PBH abundance.

In Fig.~\ref{fig:fpbh}, we plot the PBH fractional abundance, $\fpbh(\Mpbh)$, as a function of the PBH
mass $\Mpbh$ for the benchmark points listed in Tab.~\ref{tab:param_est}. For a fixed reheating
temperature, the present-day PBH abundance is extremely sensitive to the magnetic spectral index
$\nb$. We find that PBHs can be produced within specific mass windows that allow them to account for
the entire cold dark matter (CDM) abundance. In particular, this occurs for a narrow range of
reheating temperatures, $3\times10^{4}~\GeV \lesssim \Tre \lesssim 10^{5}~\GeV $.
As an explicit example, for $\Tre=2\times10^{5}\,\GeV$ and $\nb=2.674$, we obtain
$\fpbh\simeq 1$ for a corresponding PBH mass $\Mpbh\simeq 4\times10^{-16}\,M_{\odot}$. For this
specific set of parameters, the present-day magnetic field strength evaluated at the
$1~\Mpc$ scale is
$B_0(1\,\Mpc)\simeq 1.26\times10^{-22}\,\mathrm{G}$.

For other reheating temperatures, i.e., $\Tre<10^{4}\,\GeV$, although PBHs can be produced over a wider mass range, they
typically constitute only a fraction of the observed dark matter abundance. Nevertheless, these
scenarios remain consistent with the present-day magnetic field strength inferred from
observations. In particular, we find that within the mass range $10^{-12}\,M_{\odot} \leq \Mpbh \leq 10^{-1}\,M_{\odot}$,
PBHs can contribute a subdominant fraction of the present-day dark matter abundance while still
successfully explaining the observed strength of cosmic magnetic fields.

In Fig.~\ref{fig:fpbh_deltac}, we show $\fpbh$ as a function of the PBH mass for a representative set of reheating and magnetogenesis parameters, where we choose $\Tre = 10^3~\mathrm{GeV}$, $\wre = 0$, and a magnetic spectral index $\nb = 2.50$ on super-horizon scales. Here we examine the dependence of the PBH abundance on the collapse threshold $\delta_c$. Since PBHs sourced by magnetic fields experience an additional magnetic pressure compared to inflationary curvature perturbations, the effective threshold may differ. However, because the exact value of $\delta_c$ for magnetically induced perturbations is not known, we consider four representative values within the range suggested in the literature.

For a fixed reheating scenario, the peak of the magnetic power spectrum—and the associated curvature perturbation amplitude—remains unchanged, so adopting a constant $\delta_c$ is justified. Moreover, within the parameter region relevant for simultaneously explaining the present-day magnetic field strength and producing PBHs as a dark matter candidate, $\delta_c$ does not vary significantly~\cite{Kushwaha:2024zhd}.

\color{black}
\section{GWs from the Magnetic Field}\label{sec5}
In the magnetogenesis scenario considered here, a significant magnetic field is generated at small scales near $k\simeq\kre$, which can lead to PBH formation. Due to the anisotropies sourced by the magnetic field during the radiation-dominated era, we also expect a substantial production of secondary gravitational waves (SGWs). 

In this class of magnetogenesis models, the electromagnetic field can be generated both during inflation and during reheating, with its maximum energy density reached at the end of reheating. Consequently, the largest magnetic-field-induced anisotropies are expected just after the end of reheating. Beyond this point, both the magnetic field and the background radiation energy density dilute as $a^{-4}$. As shown earlier, for $\wre = 0$ (matter-like evolution during reheating), although there is continuous EM field production, the generation of SGWs from the EM field is suppressed compared to that occurring during the radiation-dominated era (see~\cite{Maiti:2025awl} for details).

Gravitational waves (GWs) can also be produced from vacuum fluctuations of spacetime during inflation, when the corresponding modes exit the horizon. This type of tensor fluctuation generation is typically referred to as the primary production of GWs (PGWs), which has a significant impact on large scales. To make this work self-contained, we briefly discuss both PGWs and SGWs in this section. In the presence of the electromagnetic field, the anisotropic stress component of the energy-momentum tensor can source the tensor perturbations, denoted as $\hij(\vx,\eta)$. The equation of motion (EoM) of the tensor perturbation is governed by ~\cite{Sorbo:2011rz, Caprini:2014mja, Ito:2016fqp, Sharma:2019jtb, Okano:2020uyr, Maiti:2025cbi, Maiti:2025rkn, Maiti:2024nhv, Bhaumik:2025kuj, PhysRevLett.102.161101}
\begin{align}
    \hij''(\vx,\eta)+2\mH \hij'(\vx,\eta)-\nabla^2\hij(\vx,\eta)=\frac{2}{\Mp^2}\mathcal{P}_{ij}^{lm}T_{lm} ,
\end{align}
where $h_{ij}$ is a traceless tensor, satisfying $\partial^i h_{ij} = h^i_i = 0$. The quantity $\mathcal{H} = a'(\eta)/a(\eta)$ represents the conformal Hubble parameter. The term $\mathcal{P}^{lm}_{ij}$ is the transverse traceless projector, given by $\mathcal{P}_{ij}^{lm} = P^l_i P^m_j - P_{ij} P^{lm}/2$, where $P_{ij} = \delta_{ij} - \partial_i \partial_j/\Delta$. The term $T_{lm}$ represents the spatial part of the energy-momentum tensor of the gauge field.

We can decompose the tensor perturbation $\hij(\vx,\eta)$ in terms of their Fourier modes $\hkl(\eta)$ as
\begin{align}
    \hij(\vx,\eta)=\sum_{\lambda=\pm}\int \frac{\d^3 \vk}{(2\pi)^3}e^{\lambda}_{ij}(\vk)\hkl(\eta)\,e^{i\vk\cdot\vx},
\end{align}
where $e^{\lambda}_{ij}(\vk)$ is the polarization tensor corresponding to the mode with wave vector $\vk$. Here $\lambda$ denotes the polarization state of the GW, which mainly depends on the nature of the source. In this magnetogeneis model, we have generated a non-helical magnetic field, where both polarization modes are equally amplified from the vacuum, so both polarization states of the produced GWs will have the same contribution. In Fourier space, the equation of motion of the tensor fluctuations is~\cite{Maiti:2025rkn, Maiti:2024nhv}
\begin{align}
    {\hkl}''+2\mH{\hkl}'+k^2\hkl=\mathcal{S}(\vk,\eta) .
\end{align}
As we have already discussed, as the magnetic field gets its maximum amplified at the end of reheating, the maximum produced anisotropies should be after completion of the reheating. After reheating electric field will decay immediately due to the high electrical conductivity, and the only contribution comes from the magnetic field part. During radiation dominated era, the source term $\mathcal{S}(\eta)$ can be written as~\cite{Maiti:2024nhv,Caprini:2014mja,Caprini:2018mtu}
\begin{align}
    \mathcal{S}(\vk,\eta)=-\frac{2\,e^{ij}_\lambda(\vk)}{\Mp^2}\int \frac{\d^3\vq}{(2\pi)^3}B_i(\vq,\eta)B_j(\vk-\vq,\eta).
\end{align}
Now we can define the two-point correlation function of the tensor fluctuation associated with the tensor power spectrum as
\begin{align}
    \mPt(k,\eta)=\sum_{\lambda=\pm}\frac{k^3}{2\pi^2}\langle \hkl(\eta)h^{\lambda *}_{\vk'}(\eta)\rangle \delta^{(3)}(\vk-\vk') .
\end{align}
Note that in these scenarios, both primary and secondary tensor fluctuations are linearly polarized, which means both polarization modes have the same contribution.

\paragraph{\underline{Production of PGWs}:\\}
It is well known that the quantum fluctuation of the spacetime, amplified during inflation, imprints as a stochastic background that carries signatures of inflationary dynamics~\cite{Starobinsky:1979ty, Grishchuk:1974ny, Guzzetti:2016mkm, Haque:2021dha}. In this de-Sitter inflation approximation, with the initial Bunch-Davies initial condition, we can write the solution of the tensor fluctuation originating from the vacuum fluctuation as~\cite{Haque:2021dha}
\begin{align}
    \hk(\eta\leq\ee)\simeq \frac{\sqrt{2}}{\Mp}\frac{i\HI}{\sqrt{2k^3}}\l[ 1-\frac{ik}{\HI a(\eta)}\r]e^{-ik\eta} .
\end{align}
Here it is clear that the vacuum fluctuations are proportional to the inflationary energy scale, i.e., $\hk\propto \HI$.

After inflation, we consider that our universe evolves under the matter-like fluid with average EoS $\wre=0$. The matter-like background yields the GW evolving as, ~\cite{Maiti:2025awl, Maiti:2025cbi, Maiti:2024nhv, Chakraborty:2024rgl, Haque:2021dha}
\begin{align}
    \hk(\eta>\ee)\simeq \frac{\pi^2x^{-3/2}}{2\Gamma^2(-3/2)\Gamma(5/2)}J_{-3/2}(x)\,\hinf(\ee).
\end{align}
where $\hinf(\ee)$ is the total tensor fluctuation produced during inflation, defined at the end of inflation. After reheating, the standard radiation-dominated era has started, and in this era, the evolution of the tensor fluctuation is simply described as~\cite{Maiti:2025awl, Maiti:2025cbi}
\begin{align}
    \hk(\eta > \eta_{\text{re}}) = x^{-1} \left( \mathcal{D}_1 e^{-ix} + \mathcal{D}_2 e^{ix} \right),
\end{align}
where \( \mathcal{D}_1 \) and \( \mathcal{D}_2 \) are constants given by
\begin{subequations}\label{eq:D}
    \begin{align}
        \mathcal{D}_1 &= \frac{\hkr(\xre) \left( i \xre - 1 \right) - \xre \hkrp(\xre)}{2i} e^{i \xre}, \\
        \mathcal{D}_2 &= \frac{\xre \hkrp(\xre) + \hkr(\xre) \left( i \xre + 1 \right)}{2i} e^{-i \xre}.
    \end{align}
\end{subequations}
Now, depending on the conditions, where the modes are outside the horizon or inside the horizon at the end of reheating, the spectral behaviour of the present-day GWs spectrum will be modified. If the modes are outside of the horizon at the end of reheating, the spectrum is typically nearly scale invariant, whereas if the modes are inside the horizon, then due to matter like evolution duration reheating, we get a red tilted spectrum. Therefore, the combined PGW spectrum is express as, as~\cite{Maiti:2025awl, Maiti:2025cbi, Maiti:2024nhv, Chakraborty:2024rgl, Haque:2021dha} 
\begin{align}
    \mPt^{\rm vac}(\eta>\ere)\simeq \frac{\mathcal{P}_{\rm T}^{\rm inf}(k,\ee)}{k^2\eta^2}
   \times \l\{
    \begin{matrix}
        1 & k\leq \kre\\
        \l(\frac{k}{\kre}\r)^{-2} & \kre<k<\ke .
    \end{matrix}
    \r. 
\end{align}

\paragraph{\underline{Production of SGWs}:}
After reheating, there is no further production of the magnetic field, as the gauge field restores its conformal nature, and due to the high electrical conductivity, the electric field decays immediately from the universe. So during a radiation-dominated era, the only existing field is the magnetic field, which is responsible for the generation of the secondary gravitational waves. However, the source due to the magnetic field only effectively produces the SGWs up to the neutrino decoupling era, i.e., $\eta=\eta_\nu$, because after neutrino decoupling, they freely travel with very high kinetic energy and consequently dilute the anisotropic stress originated due to the magnetic field~\cite{PhysRevD.70.043011, KAGRA:2021kbb}, and suppress SGW production. Therefore, the production of the secondary gravitational waves due to the magnetic field at the time of neutrino decoupling is~\cite{Maiti:2025awl, Maiti:2025cbi, Maiti:2024nhv}
\begin{align}
    \mPts&(k,\eta_\nu)=\frac{2}{\Mp^4k^4}\times\l(\int_{\xre}^{x_\nu}\d x_1 \frac{\mGk(x_\nu,x_1}{a^2(x_1)}\r)^2\nn\\
    &\times\int_0^{\infty}\frac{\d q}{q}\int_{-1}^1\d \gamma \frac{f(\gamma,\mu)\tmPb(q,\ere)\tmPb(|\vk-\vq|,\ere)}{\l[1+\l(\frac{q}{k}\r)^2-2\gamma\frac{q}{k}\r]^{3/2}},
\end{align}
where we define the Green's function during the radiation-dominated era as~\cite{Maiti:2024nhv}
\begin{align}
    \mGk(x_\nu,x_1)=\theta(\eta_\nu-\eta_1)
\frac{\eta_1}{\eta_\nu}\sin[k(\eta_1-\eta_\nu)]
\end{align}
Here $\tmPb(k,\ere)=a^4(\eta)\mPb(k,\ere)$ is the comoving magnetic power spectrum during the radiation-dominated era. Now we can write the comoving magnetic power spectrum during radiation radiation-dominant era as~\cite{Maiti:2025awl}
\begin{align}\label{eq:mPtb_s}
    \tmPb(k,\ere)=\frac{k^4|\beta|^2}{8\pi}\l(\frac{k}{\ke}\r)^{-2(\alpha+n+1)}\l(\frac{k}{\kre}\r)J^2_{\alpha+1/2}\l(\frac{k}{\kre}\r)
\end{align}
Here we deined we define $|\beta|^2=2^{2\alpha+1}\Gamma^2(\alpha+1/2)\qnb^1(n)$, where $\qnb^1$ is defined in Eq.\eqref{eq:fnb1}. Now utilizing this in the above Eq.\eqref{eq:mPtb_s}, we have found that the tensor power spectrum associated with the magnetic field at the time of the  neutrino decoupling era as~\cite{Maiti:2025awl}
\begin{widetext}
    \begin{align}
        \mPts(k,\eta_\nu)&=\frac{k^4|\mB|^4}{32\pi^2\Mp^4}\left( \frac{k}{\ke} \right)^{-4(n+\alpha+1)}\xre^2 \left( \int_{\xre}^{x_{\nu}} dx_1 \frac{\mGk(x_\nu,x_1)}{a^2(x_1)}\right)^2\nn\\
&\times \int_{u_{\rm min}}^{u_{\rm max}} \frac{du}{u}\int_{-1}^1 d\mu \frac{f(\mu,\gamma)}{[1+u^2-2u\mu]^{3/2}}u^{3-2(n+\alpha)}(|1-u|)^{3-2(n+\alpha)}\mJ^2_{\alpha+1/2}(u\xre)\mJ^2_{\alpha+1/2}(|1-u|\xre),
    \end{align}
\end{widetext}
where we defined $u=q/k$ as a new dimensionless variable$u_{\rm min}=\kpv/k$ and $u_{\rm max}=\ke/k$. Here $f(\gamma,\mu) = (1+\gamma^2)(1+\mu^2)$, with $\gamma = \hat{\mathbf{k}} \cdot \hat{\mathbf{q}}$ and $\mu = \widehat{\mathbf{k}-\mathbf{q}} \cdot \hat{\mathbf{k}}$~\cite{Sharma:2019jtb}.
To solve the above integral, we have to perform a numerical integration of the above for the entire range of $u$ (for details, please see~\cite{Maiti:2025awl}).

\paragraph{\underline{Present-day GW Spectral Energy Density $\ogwh$:}}
Gravitational waves (GWs) weakly interact with matter and can freely propagate without losing their original nature after being produced during a radiation-dominated era. The GW energy density, which scales as $\rho_{\text{GW}} \propto a^{-4}$, can be normalized by the total energy density at the production time, $\rho_c(\eta)$. Presently, this density parameter is given by
\begin{align}
    \ogw(k, \eta) = \frac{\rho_{\text{GW}}(k, \eta)}{\rho_c(\eta)} = \frac{1}{12} \frac{k^2 \mPt(k, \eta)}{a^2(\eta) H^2(\eta)},
\end{align}
where $\rho_c(\eta) = 3 H^2(\eta) \Mp^2$, with the reduced Planck mass $\Mp \approx 2.43 \times 10^{18} \, \GeV$.

The GW energy density follows the radiation scaling behavior, making modes within the Hubble radius near radiation-matter equality particularly relevant. The present-day energy density parameter $\ogw(k) h^2$ is then given by
\begin{align}
    \ogwh(k) \simeq \left(\frac{\gsp}{\gseq}\right)^{1/3} \Omega_r h^2 \ogw(k, \eta),
\end{align}
where $\Omega_r h^2 = 4.3 \times 10^{-5}$, $\gseq \simeq \gsp = 3.35$, and denote relativistic degrees of freedom at equality and today, respectively.

The spectral energy density of gravitational waves (GWs) originating from magnetic fields and from vacuum fluctuations can be distinguished by their distinct characteristics. Depending on whether a given wavenumber is super-horizon or sub-horizon at the end of reheating, the resulting spectral shape will be modified accordingly.

The behavior of primary GWs in a non-instantaneous reheating scenario is well understood. In our case, we consider a matter-like evolution after the end of inflation, with $\wre = 0$. For modes that re-enter the horizon before the end of reheating, the GW amplitude is diluted due to the background expansion, resulting in a red-tilted spectrum with $\ogwp(f > \fre) \propto {f}^{-2}$. Here, $f = 2\pi/k$. On the other hand, modes that remain well outside the horizon at the end of reheating exhibit a scale-invariant spectrum, since the tensor power spectrum generated during inflation is itself scale-invariant, originating from vacuum fluctuations in a de Sitter background. Therefore, the PGW spectrum can be expressed as~\cite{Maiti:2025awl, Maiti:2025cbi, Maiti:2024nhv, Chakraborty:2024rgl, Haque:2021dha}
\begin{align}
\ogwp h^2(f)\simeq \frac{\Omega_{\rm r}h^2}{6}\frac{\HI^2}{\Mp^2}\times\l\{
\begin{matrix}
    1 & f<\fre\\
    \mD\l(\frac{f}{\fre}\r)^{-2} & \fre<f<\fe
\end{matrix}
\r.
\end{align}
where $\mD\simeq 6/\pi^{1/2}\simeq\mathcal{O}(1) $~\cite{Maiti:2025cbi, Maiti:2024nhv}.

Similarly, the SGWs sourced by the magnetic field exhibit a broken power-law spectrum, reflecting the broken power-law nature of the magnetic field itself. In analogy to the PGW case, the spectral behavior of SGWs generated by the magnetic field can be expressed as~\cite{Maiti:2025awl}  
   \begin{align}\label{eq:sgw_mag}
 \ogws& h^2 (f)\simeq \frac{\Omega_{\rm r}h^2}{6}\l(\frac{\HI}{\Mp}\r)^4\l(\frac{\ae}{\are}\r)^2\mathcal{I}^2(\ere,\eta_\nu)\nn\\
& \times \l\{
    \begin{matrix}
       \mathcal{A}_1\l(\frac{\fe}{\fre}\r)^{4(\alpha+1)}\l(\frac{f}{\fe}\r)^{2\nb} & f_{\rm cmb}<f<\fre\\
       \mathcal{A}_2 \l(\frac{\fe}{\fre}\r)^{2{\nb}_{,2}}\l(\frac{f}{\fe}\r)^{-2{\nb}_{,2}} & \fre<f<\fe
    \end{matrix}
    \r.
\end{align} 
Here we define ${\nb}_{,2} = 2(\alpha + 1)-\nb$ with
\begin{subequations}
    \begin{align}\label{eq:amplitude}
        \mathcal{A}_1 &=\frac{|\mB|^42^{-4(n+1/2)}}{144\,\pi^2(4-2n)\Gamma^4(\alpha+1/2)},\\
        \mathcal{A}_2 &=\frac{|\mB|^4}{360\,\pi^4}\frac{1}{2(n+\alpha)-2},
    \end{align}
\end{subequations}
Here, $\mathcal{I}(\eta_\nu,\ere)$ is defined as~\cite{Maiti:2024nhv}
\begin{align}
\mathcal{I(}\ere,\eta_\nu)=\int_{\xre}^{x_\nu}dx_1\frac{\sin(x_1 - x_\nu)}{x_1} .
\end{align}
It is evident that the GW spectral index is simply twice the magnetic spectral index when the GWs are sourced by the magnetic field. This follows from the relation $\ogws(k) \propto \tmPb^2(k)$, with the dominant production occurring during the radiation-dominated era. The only additional scale dependence arises logarithmically from the factor $\mathcal{I}^2(\eta_\nu, \ere)$, since both the background and source energy densities dilute as $a^{-4}$.
\begin{figure}[t]
    \centering
    \includegraphics[width=1\linewidth]{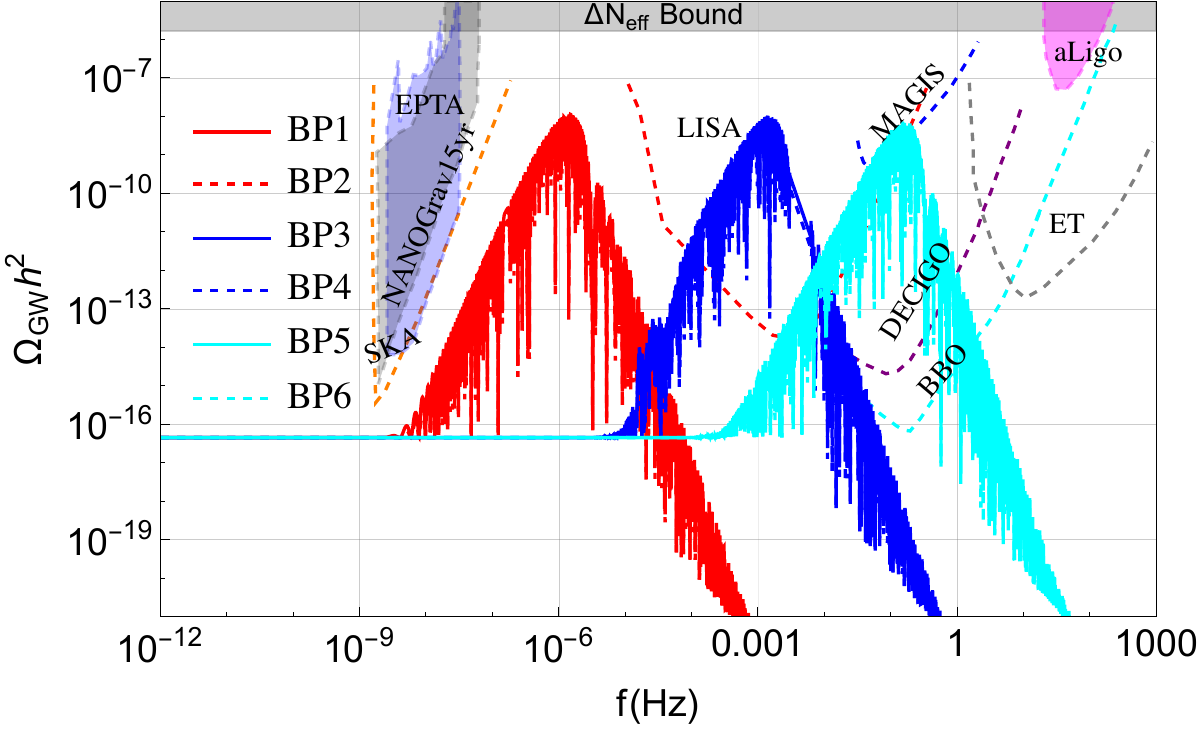}
    \caption{In this figure, we present the present-day spectral energy density of GWs, $\ogwh(f)$ sourced by the magnetic field, as a function of the observable frequency $f~(\mathrm{Hz})$ for six benchmark points (BPs) listed in Tab.~\ref{tab:param_est}. }
    \label{fig:gws}
\end{figure} 
In Fig.~\ref{fig:gws}, we present the present-day spectral energy density $\ogwh(f)$ of GWs as a function of comoving frequency $f~(\mathrm{Hz})$ for six benchmark points (BPs) listed in Tab.~\ref{tab:param_est}. For BP1 (red solid line) and BP2 (red dashed line), the GW spectra exhibit nearly identical amplitudes and shapes. This similarity arises because, for a fixed reheating temperature, a small change in the magnetic spectral index does not significantly alter the magnetic field strength, and consequently, the corresponding GW amplitude remains almost unchanged. However, as shown in Fig.~\ref{fig:fpbh}, the PBH abundance is highly sensitive to the magnetic spectral index. A similar trend is observed for the other benchmark points.

In Fig.~\ref{fig:gws}, the three different colors correspond to three distinct reheating temperatures,
$\Tre = 1~\GeV$ (red), $\Tre = 10^3~\GeV$ (blue), and $\Tre = 10^5~\GeV$ (cyan), as listed in
Tab.~\ref{tab:param_est}. Since the GW spectral energy density scales as the fourth power of the
magnetic field amplitude, $\Omega_{\rm GW} \propto B^4$, a small variation in the magnetic spectral
index $\nb$ does not lead to a significant change in the magnetic field strength
(see Eq.~\eqref{eq:amplitude}). As discussed earlier, the magnetic field amplitude is sensitive to the
magnetogenesis parameter $n$; however, small variations in its value do not significantly modify the
resulting magnetic field strength (see Tab.~\ref{tab:param_est}). Consequently, the corresponding
change in the GW spectrum, which depends on the magnetic field amplitude, is relatively mild, in
contrast to the PBH abundance, which exhibits a much stronger sensitivity, as also evident from
Tab.~\ref{tab:param_est}.

As an explicit example, for $\Tre=1\,\GeV$ and $\nb=2.28$, we find that the GW amplitude at the peak
frequency is $\ogwh(f_{\rm peak})\simeq 2.05\times10^{-9}$, whereas for a slightly larger value
$\nb=2.285$, the peak amplitude decreases to
$\ogwh(f_{\rm peak})\simeq 9.92\times10^{-10}$. In contrast, for the same set of parameters, the
present-day dark matter abundance shows a dramatic change: $\fpbh(f_{\rm peak})\simeq
7.56\times10^{-4}$ for $\nb=2.28$, while for $\nb=2.285$ it is reduced to
$\fpbh(f_{\rm peak})\simeq 5.9\times10^{-10}$.

The peak frequency of the induced GW spectrum depends sensitively on the reheating temperature,
as the peak of the magnetic power spectrum is reheating-temperature dependent. Lower reheating
temperatures correspond to a larger horizon size at the end of reheating, which in turn shifts the GW
peak toward lower frequencies. Thus, decreasing $\Tre$ moves the peak of the GW spectrum to smaller
frequency bands. For the benchmark parameter sets considered here, the peak amplitudes of the GW
spectrum remain nearly identical. This is because achieving PBH formation requires a comparable
magnetic energy density perturbation, typically $\delta\rho_B/\rho \simeq 0.1$, across the parameter
space. However, the initial PBH abundance depends exponentially on the peak amplitude of the
magnetically induced curvature perturbations, making the PBH fraction highly sensitive to even a small
shift in $\nb$.

Therefore, if PBHs are sourced by primordial magnetic fields, an associated GW background is
always generated alongside them. Importantly, the spectral shape of these magnetically induced GWs~\cite{Caprini:2018mtu, Caprini:2014mja, Maiti:2025awl, Maiti:2024nhv} is
qualitatively different from that of GWs produced by curvature perturbations, offering a distinct
observational signature~\cite{PhysRevLett.102.161101, Di:2017ndc, PhysRevD.103.083510}.

\paragraph{\underline{Comparison with scalar-induced GWs}:}
In this scenario, primordial magnetic fields source gravitational waves (GWs) through two distinct mechanisms. First, the anisotropic stress of the magnetic field directly sources tensor perturbations. Second, the curvature perturbations induced by the magnetic field act as a secondary source for tensor modes, giving rise to scalar-induced gravitational waves (SIGWs). Since these two mechanisms are physically distinct, the resulting GW spectra exhibit different behaviors across frequency ranges.

In this work, we focused so far was on the magnetic field induced GW production. The scalar perturbation should also induce the gravitational wave. Due to its secondary nature with respect to the magnetic spectrum, its contribution to GW is expected to be sub-dominant. For comparison, however we provide an analytical estimate of the scalar-induced GWs, following Refs.~\cite{Domenech:2021ztg, Atal:2021jyo}. A precise determination of the GW spectrum over the full frequency range, including its amplitude and detailed spectral features, requires a full numerical solution of the tensor perturbation equations, as performed in Refs.~\cite{Khlopov:1980mg, Bullock:1996at, Garcia-Bellido:1996mdl, Yokoyama:1998pt, Kawasaki:1997ju, PhysRevLett.102.161101, Garcia-Bellido:2017mdw, PhysRevD.103.083510, Bhaumik:2020dor, Solbi:2021wbo, Figueroa:2021zah, Frolovsky:2022qpg}.

For our present purpose of comparison, we restrict our to the spectral range $k_{\rm SB,1}<k<k_{\rm SB,2}$, where the secondary curvature perturbations dominate over the primary one. Within this window, the curvature power spectrum can be parameterized as~\cite{Domenech:2021ztg, Atal:2021jyo}
\begin{align}
    \mPc(k)\simeq \mathcal{A}_{\zeta}\times
    \l\{
    \begin{matrix}
        (k/\kre)^{n_{\rm IR}} & k<\kre\\
        (k/\kre)^{-n_{\rm UV}} & k>\kre
    \end{matrix}
    \r.
\end{align}
Here, $n_{\rm IR}=2\nb$ for $\nb\leq 3/2$ where for $\nb>3/2$ it take $n_{\rm IR}=3$, whereas the characterizes the infrared spectral tilt near the peak ($k<\kre$), while $n_{\rm UV}=2(\alpha+1)-\nb$ denotes the ultraviolet spectral tilt ($k>\kre$). The peak amplitude $\mathcal{A}_{\zeta}$ is defined in Eq.~\eqref{eq:curv_peak}.

We focus on the region of parameter space in which the model can simultaneously explain the observed large-scale magnetic field strength and generate a sufficient abundance of primordial black holes (PBHs) in the mass range compatible with cold dark matter. In this regime, we consistently find $n_{\rm IR}>3/2$ and $n_{\rm UV}<4$. As an illustrative example, for the benchmark point BP1 with $\nb=2.28$, $\wre=0$, and $\Tre=1\,\GeV$, we obtain $n_{\rm IR}=4.56$ and $n_{\rm UV}=3.058$. For this class of parameters, the GW spectrum sourced by curvature perturbations can be written as~\cite{Domenech:2021ztg, Atal:2021jyo}
\begin{align}\label{eq:sgw_cur}
    \Omega_{\rm gw}^{\rm cur}h^2\propto \mathcal{A}_{\zeta}^2\times
    \l\{
    \begin{matrix}
        (f/\fre)^3 & f<\fre\\
        (f/\fre)^{-4(\alpha+1)+2\nb} & f>\fre
    \end{matrix}
    \r.
\end{align}
Here, $\Omega_{\rm gw}^{\rm cur}h^2$ denotes the present-day spectral energy density of gravitational waves sourced by the curvature power spectrum.

Comparing Eq.\eqref{eq:sgw_mag} with Eq.\eqref{eq:sgw_cur}, we find that for modes outside the horizon at the end of reheating ($f<\fre$), the GW spectrum sourced by the magnetic field scales as $\ogws(f)\propto f^{2\nb}$, whereas the scalar-induced GW spectrum behaves as $\Omega_{\rm gw}^{\rm cur}\propto f^{3}$. For modes inside the horizon ($f>\fre$), both spectra follow the same scaling, $\Omega_{\rm gw}^{\rm sec}\propto f^{-4(\alpha+1)+2\nb}$. The peak amplitude of the GW spectrum originating from the magnetic field is larger than the contribution arising from scalar perturbations. Near the peak, the ratio of the peak amplitudes due to the magnetic field and the scalar perturbations can be estimated as $\ogw^{\rm mag}(\fre)/\ogw^{\rm cur}(\fre)\sim \mathcal{O}(10^2)$.

\color{Black}

\section{Conclusion}\label{sec6}
A wide range of mechanisms for primordial black hole (PBH) formation have been proposed in the literature, including enhancements in inflationary perturbations~\cite{Garcia-Bellido:2017mdw, Motohashi:2017kbs, Byrnes:2018txb, Ballesteros:2018wlw, Raveendran:2022dtb, Ragavendra:2020sop, 
  Garcia-Bellido:1996mdl, Garcia-Bellido:2017mdw, PhysRevD.103.083510, Solbi:2021wbo, Figueroa:2021zah, Frolovsky:2022qpg, PhysRevLett.102.161101}, bubble collisions in first-order phase transitions~\cite{PhysRevD.109.123030, PhysRevD.26.2681, Rubin:2001yw, Ai:2024cka}, cosmic-string collapse~\cite{Hawking:1987bn, Polnarev:1988dh, PhysRevD.45.3447, Balaji:2025tun, Balaji:2024rvo, Balaji:2022rsy}, and scalar-field fragmentation~\cite{PhysRevD.98.083513, Cotner:2019ykd}. However, none of these scenarios currently enjoys direct observational support regarding the true origin of PBHs. In this work, we have explored an alternative and well-motivated pathway for PBH formation in a mass window compatible with PBHs constituting the cold dark matter (CDM).

In our framework, primordial magnetic fields are generated through a sawtooth-type coupling, which naturally yields a broken power-law magnetic power spectrum while avoiding the usual backreaction and strong-coupling issues. This structure imprints a corresponding broken power-law feature in the induced curvature perturbation spectrum, with the peak scale determined by the reheating dynamics. We have shown that, for a suitable choice of model parameters and reheating temperature, the same magnetogenesis setup can both account for the observed large-scale magnetic fields and produce sufficiently enhanced density fluctuations to generate PBHs.

Since PBH formation is exponentially sensitive to the peak amplitude of curvature perturbations, and in this scenario directly linked to the peak amplitude of the primordial magnetic field, the model naturally predicts a sizable stochastic gravitational-wave (GW) background sourced by magnetic anisotropic stresses. The spectral shape of these secondary GWs is governed by the magnetic spectral index, making it qualitatively distinct from the spectrum produced by curvature perturbations~\cite{Khlopov:1980mg, Bullock:1996at, Garcia-Bellido:1996mdl, Yokoyama:1998pt, Kawasaki:1997ju, PhysRevLett.102.161101, Garcia-Bellido:2017mdw, PhysRevD.103.083510, Bhaumik:2020dor, Solbi:2021wbo, Figueroa:2021zah, Frolovsky:2022qpg}. For appropriate parameter choices, the predicted GW signals fall within the target sensitivity of future experiments such as LISA~\cite{Amaro-Seoane:2012aqc, Barausse:2020rsu}, DECIGO~\cite{Seto:2001qf, Kawamura:2011zz, Suemasa:2017ppd}, BBO~\cite{Crowder:2005nr, Corbin:2005ny, Baker:2019pnp}, and SKA~\cite{Janssen:2014dka}, and may also help explain the PTA signals reported in recent observations~\cite{NANOGrav:2023gor, 2023arXiv230616224A, Reardon:2023gzh, Zic:2023gta, Xu:2023wog}.

Furthermore, both the PBH mass function and the peak frequency of the induced GW spectrum depend sensitively on the reheating temperature. Thus, a joint observation of a characteristic GW peak together with a PBH mass range compatible with CDM could provide key information about reheating and the underlying magnetic power spectrum. Multi-messenger constraints—combining PBH abundance, GW observations, and large-scale magnetic-field measurements—are therefore essential to fully test this class of magnetogenesis scenarios.

In summary, this work demonstrates that a unified magnetogenesis framework can simultaneously generate observable primordial magnetic fields, produce distinctive gravitational-wave signatures, and provide a viable mechanism for PBH formation. This connection between early-Universe magnetic fields and present-day dark matter abundance offers a promising target for upcoming observational probes.

 \acknowledgments
SM gratefully acknowledges financial support from the Council of Scientific and Industrial Research (CSIR), Ministry of Science and Technology, Government of India.

\appendix

\bibliographystyle{apsrev4-1}
\bibliography{references}

\end{document}